\documentclass{article}

\usepackage[affil-it]{authblk}
\usepackage[usenames,dvipsnames]{xcolor}
\usepackage{amsfonts}
\usepackage{amsmath,amsthm,amssymb,dsfont}
\usepackage{enumerate}
\usepackage[english]{babel}
\usepackage{graphicx}	
\usepackage[caption=false]{subfig}
\usepackage[margin=3cm]{geometry}
\usepackage{url}
\usepackage{todonotes}
\usepackage{bbm}

\usepackage{tikz}
\usetikzlibrary{chains}
\usetikzlibrary{fit}

\usepackage{makecell}

\usepackage{epsfig}
\usetikzlibrary{shapes.symbols,patterns} 
\usepackage{pgfplots}

\usepackage{hyperref}
\hypersetup{colorlinks=true,citecolor=blue,linkcolor=blue,filecolor=blue,urlcolor=blue,breaklinks=true}

\usepackage{nicefrac}
\usepackage{mathtools}

\usepackage{wasysym}

\usepackage{booktabs}

\usepackage{braket}
\usepackage{cleveref}
\usepackage{qcircuit}
\usepackage{hyphenat}
\usepackage[ruled,vlined]{algorithm2e}
\usepackage{bbold}
\usepackage{afterpage}

\theoremstyle{plain}
\newtheorem{theorem}{Theorem}[section]

\theoremstyle{definition}

\newtheorem{assumption}[theorem]{Assumption}

\newcommand*{\CNOT}{\mathrm{CNOT}}
\newcommand*{\SWAP}{\mathrm{SWAP}}
\newcommand*{\Ry}{\mathrm{Ry}}

\newcommand*{\cE}{\mathcal{E}}
\newcommand*{\cF}{\mathcal{F}}

\newcommand*{\cN}{\mathcal{N}}

\newcommand*{\cU}{\mathcal{U}}

\newcommand*{\X}{\mathbb{X}}

\newcommand*{\R}{\mathbb{R}}

\newcommand*{\rank}{\mathrm{rank}}

\newcommand*{\tr}{\mathrm{tr}}

\newcommand{\proj}[1]{|#1\rangle\!\langle #1|}

\newcommand*{\TPCP}{\mathrm{TPCP}}
\newcommand*{\TP}{\mathrm{TP}}

\DeclareSymbolFont{mymathoperators}{OT1}{phv}{m}{n}
\DeclareMathSymbol{\protoast}{\mathbin}{mymathoperators}{"2A}

\newcommand{\norm}[1]{\left\lVert#1\right\rVert}

\newcommand{\iu}{\mathrm{i}}
\newcommand{\sgn}{\mathrm{sgn}}

\newcommand{\linops}{\mathrm{L}}
\newcommand{\unitaryops}{\mathrm{U}}

\DeclarePairedDelimiter\abs{\lvert}{\rvert}
\DeclarePairedDelimiter\ceil{\lceil}{\rceil}

\renewcommand{\epsilon}{\ensuremath\varepsilon}

\renewcommand{\phi}{\ensuremath{\varphi}}

\SetKwBlock{Repeat}{repeat}{}

\crefname{equation}{}{}  

\allowdisplaybreaks

\title{Quasiprobability decompositions with reduced\\ sampling overhead}

 \author{\normalsize Christophe Piveteau$^{1,2}$, David Sutter$^{1}$, and Stefan Woerner$^{1}$}
 \affil{\small $^{1}$IBM Quantum, IBM Research -- Zurich\\
 \small $^{2}$Institute for Theoretical Physics, ETH Zurich\\
 }
 \date{}

\begin{document}

\maketitle

\begin{abstract}
Quantum error mitigation techniques can reduce noise on current quantum hardware without the need for fault-tolerant quantum error correction. For instance, the quasiprobability method simulates a noise-free quantum computer using a noisy one, with the caveat of only producing the correct expected values of observables. The cost of this error mitigation technique manifests as a sampling overhead which scales exponentially in the number of corrected gates. In this work, we present a new algorithm based on mathematical optimization that aims to choose the quasiprobability decomposition in a noise-aware manner. This directly leads to a significantly lower basis of the sampling overhead compared to existing approaches. A key element of the novel algorithm is a robust quasiprobability method that allows for a tradeoff between an approximation error and the sampling overhead via semidefinite programming.

\end{abstract}

\section{Introduction}
Quantum computing holds the promise for significant advancements in various fields such as quantum chemistry, material science, optimization, and machine learning. 
Unlike classical computers, quantum computers suffer from a significant amount of noise which accumulates over the execution of a quantum algorithm and thus limits experiments to shallow quantum circuits. 
The theory of quantum error correction and fault-tolerant quantum computing solves this problem. 
The celebrated \emph{threshold theorem} ensures that given the noise level of the physical gates is below some \emph{constant} threshold value, arbitrarily long calculations are possible at arbitrarily low error rates~\cite{AB08}.
The cost for a fault-tolerant implementation of a given circuit is a polylogarithmic overhead that is tame when speaking of asymptotics, but currently available quantum hardware does not fulfil the requirements for quantum error correction~\cite{Preskill2018} and it remains a major challenge to achieve this goal.

It is thus a natural question if it is possible to mitigate noise on quantum hardware without the need of full quantum error correction. Multiple schemes have been proposed~\cite{McClean2017,Li2017,Temme2017,Otten2018} that aim to reduce the effect of noise while also being significantly easier to implement than quantum error correction. All of these methods come with a certain drawback that prohibits them from achieving large-scale fault tolerant quantum computation. The term \emph{quantum error mitigation} (QEM) is often used to group these methods. The hope is to use error mitigation techniques to demonstrate a quantum advantage on a useful task with near-term devices.

One specific method in the family of quantum error mitigation techniques is the \emph{quasiprobability method}~\cite{Temme2017} (also known as \emph{probabilistic error cancellation}). The central idea is to decompose the ideal (noise-free) quantum circuit into a quasiprobabilistic mixture of noisy circuits, called \emph{quasiprobability decomposition} (QPD), that one can implement on a given hardware. These noisy circuits are randomly sampled and run. By correctly post-processing the measurement outcome, one can obtain an unbiased estimator for the expectation values of the ideal circuit. The quasiprobability method exhibits a simulation overhead in terms of an additional sampling cost. More precisely, the number of shots required to simulate a circuit scales as \smash{$\mathcal{O}(\gamma^{2|C|})$} where $\gamma\geq 1$ is the so-called \emph{sampling overhead} of the QPD and $|C|$ denotes the number of corrected gates in a given circuit $C$. This quantity encapsulates how strongly the method has to compensate for the noise in the quantum system. Notably, $\gamma$ goes to $1$ in the limit where the noise vanishes, making the simulation overhead disappear. This exponential cost restricts the quasiprobability method to shallow quantum circuits.

By the above reasoning, it is evident that one wants to find QPDs that exhibit the smallest possible $\gamma$-factor. The arguably most difficult part of finding a suitable QPD is how to choose the noisy quantum circuits, called \emph{decomposition set}, into which we decompose the ideal quantum circuit. It has been realized~\cite{Temme2017} that the optimal QPD can be expressed as a linear program, under the assumption that the decomposition set is already fixed. Furthermore, a concrete decomposition set was proposed in~\cite{Endo2018} that suffices to decompose any circuit under the constraint of arbitrary noise that is not too strong. However, this decomposition set does not necessarily produce the optimal QPD and indeed we demonstrate that there exist decomposition sets that yield a lower sampling overhead. This motivated the study of improved decomposition sets for the quasiprobability method. In~\cite{takagi20} the authors prove analytical bounds on the best possible sampling overhead under the assumption that all unitary and state preparation operations exhibit the same noise.

In this work we introduce a novel method, which we call \emph{Stinespring} algorithm, for finding efficient decomposition sets for single-qubit and two-qubit gates\footnote{Any quantum circuit can be decomposed into single-qubit and two-qubit gates, therefore it is sufficient to obtain quasiprobability decompositions of these gates in order to error mitigate an arbitrary quantum circuit.}. Based on mathematical optimization techniques for convex and non-convex problems this new iterative algorithm takes into account the hardware noise. As a building block for the Stinespring algorithm, we introduce a robust quasiprobability method called \emph{approximate quasiprobability decomposition}, which is interesting on its own. Instead of perfectly simulating a certain circuit, we allow for a small approximation error. This enables us to make a tradeoff between the approximation quality and the sampling overhead of the quasiprobability method. We will illustrate our results with simulations showing that the new methods significantly reduce the $\gamma$-factor of the QPD compared to existing approaches.

\section{Quasiprobability method}
Consider a linear operator $\mathcal{F}\in \linops(\linops(A),\linops(B))$, where $\linops(A,B)$ denotes the set of linear maps from Hilbert spaces $A$ to $B$ and $\linops(A)\coloneqq\linops(A,A)$. 
A QPD of $\mathcal{F}$ is a finite set of tuples $\{(a_i,\mathcal{E}_i)\}_i$ with $a_i\in\R$ and completely positive maps $\mathcal{E}_i$, such that for all $\rho\in \linops(A)$
  \begin{align}\label{eq:qpd}
    \mathcal{F}(\rho) = \sum\limits_i{a_i \mathcal{E}_i(\rho)}\, .
  \end{align}
We call $\{\mathcal{E}_i\}_i$ the decomposition set of the QPD. 
The quasiprobability method requires a QPD $\{(a_i,\mathcal{E}_i)\}_i$ where the maps $\mathcal{E}_i$ correspond to operations that can be implemented on the quantum hardware, e.g.~they might correspond to the channel representing a noisy quantum gate. In practice, these could be obtained by doing tomography and/or by using prior knowledge of the experimental noise. Note that the $\mathcal{E}_i$ may not need to be trace-preserving as non-trace-preserving maps can be simulated using measurements and postselection.

The quasiprobability method allows us to simulate a noiseless execution of $\mathcal{F}$ using the noisy operations $\mathcal{E}_i$. In contrast to quantum error correction, this technique is very hardware-friendly as we do not encode our quantum information in a larger space, so a logical qubit still corresponds to a physical qubit.
However, the quasiprobability method is only able to systematically remove the bias from the expectation values and cannot extend the coherence times of the computation. This manifests in a bad asymptotic scaling emphasizing that this technique is most favorable for near-term circuits.

Typically one is interested in the case where $\mathcal{F}$ is a unitary operation, i.e.~$\mathcal{F}=\mathcal{U}$ for some unitary matrix $U$, where we use the notation $\mathcal{U}(\rho)\vcentcolon = U\rho U^{\dagger}$. This corresponds to simulating an ideal quantum gate $U$ using noisy operations. But sometimes it can also be interesting to consider $\mathcal{F}$ which are non-physical, for example $\mathcal{F}$ could be the inverse of a noise operator that occurred on a certain quantum system. By simulating the inverse, one can revert the effect of the noise on average. Typically the inverse of interesting noise models, such as depolarizing noise, are not completely positive maps. From now on we will restrict ourselves to the case $\mathcal{F}$ is unitary, though all results also hold in a more generalized setting.

\subsection{Quasiprobability sampling}\label{sec:qpsampling}
Consider a gate described by a unitary $U$ and a QPD $\{(a_i,\mathcal{E}_i)\}_i$ of the unitary channel $\mathcal{U}$. Suppose that we are interested in the expectation value of a projective measurement described by a Hermitian operator $O$, which would be performed after the gate $U$, i.e.~we would like to obtain $\tr[O\,\mathcal{U}(\rho)]$, given a certain input state $\rho$. The linearity of the trace together with~\Cref{eq:qpd} implies
\begin{equation}\label{eq:qpd_linearity}
    \tr[O\,\mathcal{U}(\rho)] = \gamma \sum_i{\frac{\abs{a_i}}{\gamma} \sgn(a_i)\tr[O\mathcal{E}_i(\rho)]} \, ,
\end{equation}
where $\sgn(\cdot)$ denotes the sign function. Via~\Cref{eq:qpd_linearity} we have introduced the $\gamma$-factor $\gamma\vcentcolon =\sum_i{\abs{a_i}}$. The right-hand side of~\Cref{eq:qpd_linearity} naturally gives us a way to construct an unbiased Monte Carlo estimator for $\tr[O\,\mathcal{U}(\rho)]$, while only having access to the operations $\mathcal{E}_i$ of the noisy hardware, as depicted in~\Cref{fig:qpd}. We choose a random number $i$ with probability $|a_i|/\gamma$ and then perform the operation $\mathcal{E}_i$ followed by the measurement $O$. The measurement outcome is weighted by the coefficient $\gamma \sgn(a_i)$ which gives the output of the estimator. By sampling this estimator many times and considering the average value, we can obtain an arbitrarily precise estimate of the true expectation value $\tr[O\, \mathcal{U}(\rho)]$.

\begin{figure}[!htb]
\centering
\begin{tikzpicture}
\node at (0,-1.25) {
\Qcircuit @C=1em @R=.7em {
&  \gate{U} & \meter &\cw
}};
\node[] at (5,0) {
\Qcircuit @C=1em @R=.7em {
&  \gate{\cE_1} & \meter &\cw
}};
\node[] at (7.25,0){$\gamma\, \sgn(a_1)$};
\node[] at (5,-1) {
\Qcircuit @C=1em @R=.7em {
&  \gate{\cE_2} & \meter &\cw
}};
\node[] at (7.25,-1){$\gamma\, \sgn(a_2)$};
\node[] at (5,-2.5) {
\Qcircuit @C=1em @R=.7em {
&  \gate{\cE_M} & \meter &\cw
}};
\node[] at (7.25,-2.5){$\gamma\, \sgn(a_M)$};
\node[rotate=90] at (5,-1.75) {\ldots};
\node[rotate=90] at (7.25,-1.75) {\ldots}; 

\draw[->] (1.25,-1.25)--(3.8,0);
\draw[->] (1.25,-1.25)--(3.8,-1);
\draw[->] (1.25,-1.25)--(3.75,-2.5);

\node[] at (2.5,0.0){$p_1 = \frac{|a_1|}{\gamma}$};
\node[] at (2.75,-1.4){$p_2 = \frac{|a_2|}{\gamma}$};
\node[] at (2,-2.3){$p_M = \frac{|a_M|}{\gamma}$};

\node[] at (4,-4){$\tr[O\, \mathcal{U}(\rho)] = \sum_{i=1}^M p_i \tr[O\, \cE_i(\rho)] \gamma\, \sgn(a_i) \quad \forall \rho$};

\end{tikzpicture}     
    \caption{Graphical representation of the quasiprobability method. The ideal quantum gate $U$ is randomly replaced by a quantum channel $\mathcal{E}_i$ that the hardware can implement. The output of the measurement $O$ must be weighted according to the sampled operation.}
    \label{fig:qpd}
\end{figure}

In practice the $\mathcal{E}_i$ are imperfect estimates of the true underlying quantum channels, produced by tomography. Unfortunately, tomography fundamentally cannot give us an arbitrarily precise estimate of the true channels, due to state preparation and measurement (SPAM) errors. An erroneous knowledge of the $\mathcal{E}_i$ will give erroneous coefficients $a_i$ in the QPD, which again translate into an error in our estimator of the expectation value. It was shown that this problem can be circumvented by using gate set tomography (GST), as it still allows us to obtain an unbiased Monte Carlo estimator~\cite{Endo2018}, even though the exact $\mathcal{E}_i$ are unknown. For simplicity, we will assume from now on that we are able to perform exact tomography of the quantum channel $\mathcal{E}_i$.

The quasiprobability estimator constructed above does generally not exhibit the identical statistics as the ideal measurement outcome of $O$, it only has the same expectation value. In fact, the variance of the estimator increases with $\gamma$ and one requires $\mathcal{O}(\gamma^2)$ more shots to estimate  $\tr[O\, \mathcal{U}(\rho)]$ to a target accuracy compared to the case where $\mathcal{U}$ is implemented exactly. This is a direct consequence of Hoeffding's inequality.

The quasiprobability method can be applied to large quantum circuits consisting of many different gates, by obtaining a QPD of each quantum gate individually, and then combining them together into one large QPD of the whole circuit. The sampling of the total quasiprobability estimator can still be done efficiently: Consider a circuit consisting of a sequence of $m$ unitary gates $\mathcal{U}_m\circ\dots\circ\mathcal{U}_1$ followed by a measurement described by the observable $O$. For each operation $\mathcal{U}_k$ a QPD \smash{$\{(a_i^{(k)}, \mathcal{E}_i^{(k)})\}$} with $\gamma$-factor $\gamma_k$ is given. Our estimator starts by sampling a random number $i_1$ according to the probabilities \smash{$\{\abs{a_i^{(1)}}/\gamma_1\}$} and executing the operation \smash{$\mathcal{E}_{i_1}^{(1)}$}. In a second step we sample a random number $i_2$ according to the probabilities \smash{$\{\abs{a_i^{(2)}}/\gamma_2\}$} and execute the operation \smash{$\mathcal{E}_{i_2}^{(2)}$}. This procedure is continued for all $i=3,\dots,m$ while keeping track of all indices $i_1,\dots,i_m$ sampled along the way. At the very end we measure the observable $O$ on the system. The estimator then outputs the outcome of that measurement multiplied by \smash{$\sgn(\prod_{k=1}^m{a_{i_k}^{(k)}})\prod_{k=1}^m{\gamma_k}$}.

We see that the combined $\gamma$-factor scales in a multiplicative way as $\gamma_{\mathrm{total}}=\prod_{k=1}^m{\gamma_k}$. Therefore the sampling overhead of the total circuit scales exponentially in the circuit size.
The Monte Carlo sampler for multiple error mitigated quantum gates only remains an unbiased estimator under the assumption that the noise is localized and Markovian. In looser terms this means that the noise on any quantum gate must be uncorrelated with other noise and independent on what operations were performed previously on the circuit. Similarly to previous works we will assume for simplicity that this assumption holds~\cite{Temme2017,Endo2018}. Some more recent research has shown that cross-correlations can be tackled by using variants of the quasiprobability method which do not rely on tomography to find the optimal quasiprobability coefficients~\cite{Armands2020}.

\subsection{Finding quasiprobability decompositions}
For a given quantum hardware, can we be certain that there exists a QPD of $\mathcal{U}$ into operations that the hardware can implement? This question may seem difficult, as its answer depends on the exact details of the capabilities of the hardware, namely what kind of quantum operations it can implement and at what fidelity it does so. Fortunately the answer to this question is positive, as long as as the noise is not too strong~\cite{Endo2018}. 

Consider the simplest case where $\mathcal{U}$ is a single-qubit gate. \Cref{tab:endobasis} lists 16 single-qubit operations that can be realized by the successive execution of the Hadamard gate $H$, the phase gate $S$, and/or a measurement\hyp{}postselection operation $P_0 =\proj{0}$. The measurement\hyp{}postselection operation can be simulated in the context of the quasiprobability sampling estimator by performing a measurement in the computational basis and aborting if the outcome is $1$, where aborting means that the estimator outputs $0$. Therefore any quantum computer able to implement Clifford gates and measurements in the computational basis can also implement these operations, at least approximately. It can be verified that the mentioned set of 16 operations forms a basis of the (real) space of 1-qubit Hermitian\hyp{}preserving operations\footnote{The space spanned by the 16 operations contains only Hermitian\hyp{}preserving maps, as the 16 basis elements are themselves Hermitian\hyp{}preserving. Every Hermitian\hyp{}preserving map has a real $16\times 16$ Pauli transfer matrix, which can be linearly decomposed into the Pauli transfer matrices of the 16 basis operations.} and hence we call it \emph{standard basis}. It can be shown that the standard basis is robust to small errors in the sense that it remains a basis of the space of Hermitian\hyp{}preserving operations if the individual elements suffer from a small amount of noise~\cite{Endo2018}.

\begin{table}[!htb]
    \centering
    \begin{tabular}{|l c l c l|}
        \hline
         $[\mathds{1}]$ & & & &   \\
         $[\sigma^X]$ & & & $=$ & $[H][S]^2[H]$ \\
         $[\sigma^Y]$ & & & $=$ & $[H][S]^2[H][S]^2$ \\
         $[\sigma^Z]$ & & & $=$ & $[S]^2$ \\
         $[R_X]$ & $=$ &$[\frac{\mathds{1}+\iu \sigma^X}{\sqrt{2}}]$ & $=$ & $[H][S]^3[H]$ \\
         $[R_Y]$ & $=$ &$[\frac{\mathds{1}+\iu \sigma^Y}{\sqrt{2}}]$ & $=$ & $[S][H][S]^3[H][S]^3$ \\
         $[R_Z]$ & $=$ &$[\frac{\mathds{1}+\iu \sigma^Z}{\sqrt{2}}]$ & $=$ & $[S]^3$ \\
         $[R_{YZ}]$ & $=$ &$[\frac{\sigma^Y+\sigma^Z}{\sqrt{2}}]$ & $=$ & $[H][S]^3[H][S]^2$ \\
         $[R_{ZX}]$ & $=$ &$[\frac{\sigma^Z+\sigma^X}{\sqrt{2}}]$ & $=$ & $[S]^3[H][S]^3[H][S]^3$ \\
         $[R_{XY}]$ & $=$ &$[\frac{\sigma^X+\sigma^Y}{\sqrt{2}}]$ & $=$ & $[H][S]^2[H][S]^3$ \\
         $[\pi_X]$ & $=$ &$[\frac{\mathds{1}+\sigma^X}{2}]$ & $=$ & $[S][H][S][H][P_0][H][S]^3[H][S]^3$ \\
         $[\pi_Y]$ & $=$ &$[\frac{\mathds{1}+\sigma^Y}{2}]$ & $=$ & $[H][S]^3[H][P_0][H][S][H]$ \\
         $[\pi_Z]$ & $=$ &$[\frac{\mathds{1}+\sigma^Z}{2}]$ & $=$ & $[P_0]$ \\
         $[\pi_{YZ}]$ & $=$ &$[\frac{\sigma^Y+\iu\sigma^Z}{2}]$ & $=$ & $[S][H][S][H][P_0][H][S][H][S]^3$ \\
         $[\pi_{ZX}]$ & $=$ &$[\frac{\sigma^Z+\iu\sigma^X}{2}]$ & $=$ & $[H][S]^3[H][P_0][H][S][H][S]^2$ \\
         $[\pi_{XY}]$ & $=$ &$[\frac{\sigma^X+\iu\sigma^Y}{2}]$ & $=$ & $[P_0][H][S]^2[H]$ \\
         \hline
    \end{tabular}
    \caption{16 basis operations constituting the standard basis, which was introduced in~\cite{Endo2018} using the notation $[U](\cdot)=U(\cdot)U^\dagger$. All these operations can be realized using Hadamard gates $H$, phase gates $S$ and measurement\hyp{}postselection operations $P_0$.}
    \label{tab:endobasis}
\end{table}

Consider the setting where we have a fixed decomposition set $\{\mathcal{E}_i\}_i$ and we would like to find the optimal (in terms of lowest possible $\gamma$-factor) quasiprobability coefficients $a_i$ such that~\Cref{eq:qpd} is fulfilled. If our decomposition set consists of linearly independent operations, which is e.g.~the case for the standard basis, then this problem corresponds to solving a set of linear equations. But in practice one is often confronted with the more general case where there is an infinity of possible decompositions, so we require some method to pick out the best one of them.
This optimization problem can be written in terms of a linear program (LP)
\begin{align}\label{eq:temme_lp}
\min \limits_{a_i\in\mathbb{R}}\Big \{\sum_i |a_i| : \mathcal{F} = \sum\limits_i{a_i\mathcal{E}_i} \Big \} \, ,
\end{align}
and hence can be solved efficiently~\cite{Temme2017}.
We note that the number of equality constraints in~\Cref{eq:temme_lp} is exponential in the number of qubits involved in the untiary $\mathcal{U}$. This exponential cost implies that the LP can only be solved for few-qubit gates, typically $1$- and $2$-qubit gates.

\section{Approximate quasiprobability decomposition} \label{sec_approx_QPD}
In this section we present our first result, which is a relaxation of the QPD where we allow for an error in the decomposition. In mathematical terms, we require the condition~\eqref{eq:qpd} to hold approximately, i.e.,
\begin{equation}\label{eq:aqpd}
    \mathcal{U}(\rho) \approx \sum\limits_i{a_i \mathcal{E}_i(\rho)}\, .
\end{equation}
An erroneous QPD leads to an error in the Monte Carlo estimator. Giving up the exactness of the quasiprobability method allows us to find a better decomposition in terms of the $\gamma$-factor. More precisely the approximate QPD results in a tradeoff between the sampling overhead and the error in the method. 

\subsection{Semidefinite programming relaxation using the diamond norm}
In a first step we have to quantify the error in the approximate QPD given by~\Cref{eq:aqpd}. A natural candidate is to use the diamond norm as it has a strong operational interpretation. In addition, the diamond norm fits very naturally in our mathematical optimization setting, as it has been shown to be expressible as a semidefinite program (SDP)~\cite{Watrous2013}.

\begin{theorem}[SDP for diamond norm~\cite{Watrous2013}]
    Let $\mathcal{G}\in \linops(\linops(A),\linops(B))$ and denote its Choi matrix by $\Lambda_{\mathcal{G}}$. Then
    \begin{align} \label{eq:watrousSDP}
     \norm{\mathcal{G}}_{\diamond}=   \left \lbrace
        \begin{array}{r l}
            \max \limits_{\substack{\rho_0,\rho_1\in \linops(A)\\X\in \linops(B\otimes A)}} & \frac{1}{2}\langle\Lambda_{\mathcal{G}},X\rangle + \frac{1}{2}\langle\Lambda_{\mathcal{G}}^*,X^*\rangle \\
            \textnormal{s.t.} & \begin{pmatrix}\mathds{1}_B\otimes\rho_0 & X \\ X^* & \mathds{1}_B\otimes\rho_1\end{pmatrix} \geq 0 \\
            &\rho_0\geq 0, \rho_1\geq 0,\rho_0^{\dagger}=\rho_0, \rho_1^{\dagger}=\rho_1 \, ,
        \end{array} \right .
    \end{align}
    which is a SDP.\footnote{$X^*$ and $\Lambda_\mathcal{G}^*$ denote the adjoint operators of $X$ and $\Lambda_\mathcal{G}$, respectively. }
\end{theorem}
The dual formulation of the SDP~\eqref{eq:watrousSDP} is given by
\begin{align}\label{eq:watrousSDPdual}
    \norm{\mathcal{G}}_{\diamond}=   \left \lbrace
    \begin{array}{r l}
        \min \limits_{Y_0,Y_1\in \linops(B\otimes A)} &  \frac{1}{2}\norm{\tr_B[Y_0]}_{\infty} + \frac{1}{2}\norm{\tr_B[Y_1]}_{\infty} \\
        \textnormal{s.t.} & \begin{pmatrix}Y_0 & -\Lambda_{\mathcal{G}} \\ -\Lambda_{\mathcal{G}}^* & Y_1\end{pmatrix} \geq 0 \\
        &Y_0\geq 0, Y_1\geq 0\, ,
    \end{array}
    \right .
\end{align}
where $\norm{\cdot}_{\infty}$ denotes the spectral norm.
Suppose we have a fixed decomposition set $\{\mathcal{E}_i\}_i$ and we would like to find the best possible approximate QPD of an operation $\mathcal{U}$ into that decomposition set. More precisely, we give a certain $\gamma$-factor budget, denoted $\gamma_{\textnormal{budget}}$, such that the QPD may not have a $\gamma$-factor higher than that. The corresponding optimization problem is
\begin{align} \label{eq:budgetingproblem}
\left \lbrace
\begin{array}{r l}
    \min \limits_{a_i\in\mathbb{R}} & \norm{\mathcal{U} - \sum_i a_i\mathcal{E}_i}_{\diamond} \\
    \textnormal{s.t.} & \sum\limits_i{|a_i|}\leq \gamma_{\textnormal{budget}} \, .
    \end{array} \right .
\end{align}
By inserting the SDP from~\Cref{eq:watrousSDPdual} into~\Cref{eq:budgetingproblem} the problem can be rewritten as
\begin{align}\label{eq:aqpd_sdp}
\left \lbrace
\begin{array}{r l}
    \min \limits_{a_i\in\mathbb{R},Y_0,Y_1\in \linops(A\otimes B)} & \frac{1}{2}\norm{\tr_B[Y_0]}_{\infty} + \frac{1}{2}\norm{\tr_B[Y_1]}_{\infty} \\
    \textnormal{s.t.} & \begin{pmatrix}Y_0 & \sum_i a_i\Lambda_{\mathcal{E}_i} - \Lambda_{\mathcal{U}} \\ \sum_i a_i\Lambda_{\mathcal{E}_i}^* - \Lambda_{\mathcal{U}}^* & Y_1\end{pmatrix} \geq 0 \\ 
     & \sum_i |a_i| \leq \gamma_{\textnormal{budget}} \\ 
     &Y_0\geq 0, Y_1\geq 0 \, ,
    \end{array} \right .
\end{align}
where $\Lambda_{\mathcal{U}}$ and $\Lambda_{\mathcal{E}_i}$ are the Choi matrices of $\mathcal{U}$ and $\mathcal{E}_i$, respectively. Luckily~\Cref{eq:aqpd_sdp} is still a SDP, which allows us to efficiently solve our optimization problem.
The approximation $\mathcal{U}_{\textnormal{approx}}\coloneqq\sum_i a_i\mathcal{E}_i$ of $\mathcal{U}$ obtained from the optimization problem in~\Cref{eq:budgetingproblem} is not guaranteed to be a physical map. In practice, this implies that we might simulate the execution of a non-physical map using the quasiprobability method. It is possible to enforce complete positivity and/or trace\hyp{}preservingness of $\mathcal{U}_{\textnormal{approx}}$ into the optimization problem by adding a positivity/partial-trace constraint on the Choi matrix $\Lambda_{\mathcal{U}_{\textnormal{approx}}}$ of $\mathcal{U}_{\textnormal{approx}}$.

We note that the idea of considering approximate QPDs has been considered in some form or another in other works. For instance, in~\cite{Armands2020} the quasiprobability coefficients of a complete circuit are optimized all at once in order to minimize the error of the computation. This optimisation problem is very difficult and does not come with the same properties and guarantees as a SDP. 
In~\cite{cai21} the authors utilize the quasiprobability method to simulate noise with reduced strength in order to be used in conjunction with the error extrapolation method. The used QPD is exact and not approximate, but the target channel $\mathcal{F}$ is chosen to be noisy instead of an ideal unitary channel.

\subsection{Tradeoff curves}\label{sec:tradeoff_curves}
\begin{figure}[!htb]
    \centering
    \includegraphics[scale=0.75]{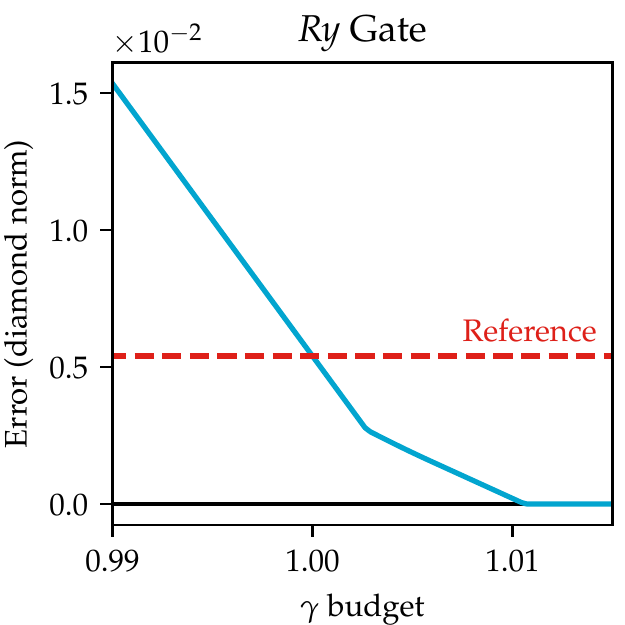} \hspace{4mm}
    \includegraphics[scale=0.75]{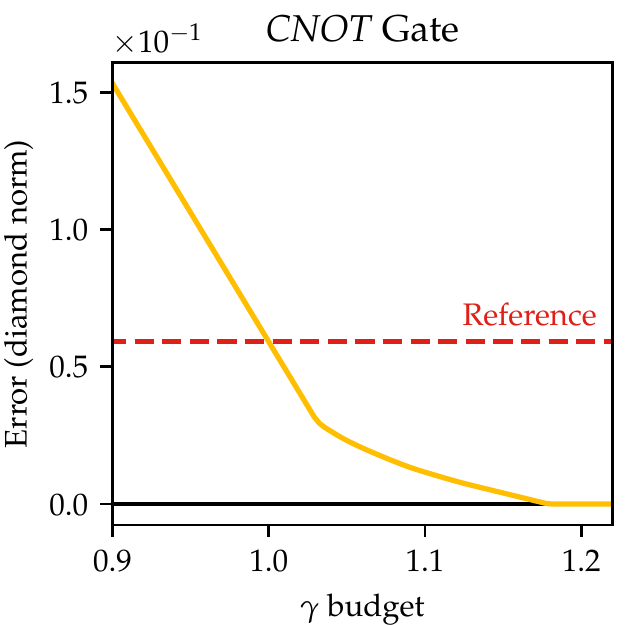}\hspace{4mm}
    \includegraphics[scale=0.75]{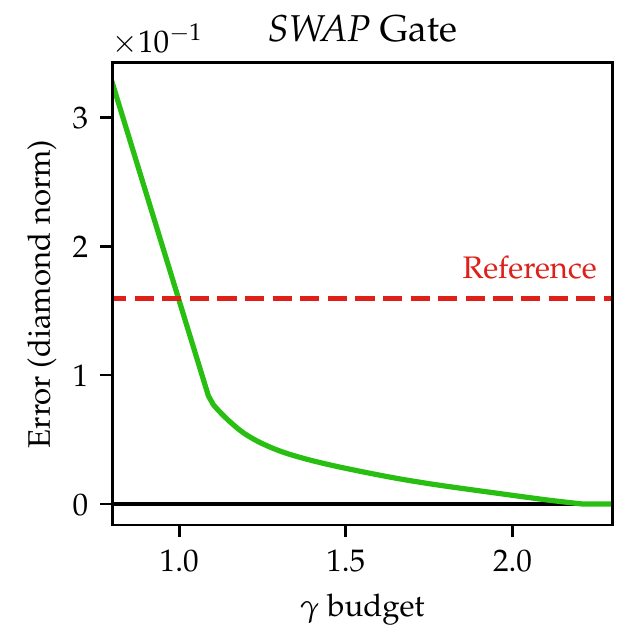}
    \caption{Tradeoff curves for the three quantum gates $\Ry$ (with an arbitrary angle since the noise model is the same for all rotation angles), $\CNOT$ and $\SWAP$ using the standard basis as decomposition set. The noisy channels of the three gates and the standard basis are extracted from a noise model in Qiskit~\cite{Qiskit} which approximates the noise on the \texttt{ibmq\_melbourne} device. More precisely, the noise model consists of a combination of depolarizing and thermal relaxation errors where the noise parameters are inferred from physical quantities measured on the hardware (e.g. $T_1,T_2$, gate times, qubit frequency, etc...). The \emph{CNOT} gates exhibit an error rate of around $2\%$ and the single-qubit gates exhibit an error rate of around $0.1\%$. The red dashed line represents the error (in diamond norm) of the reference noisy channel, i.e.~when the gate is implemented as-is without QEM. We use the SDP solver in the MOSEK~\cite{Mosek} software package through the CVXPY~\cite{Diamond2016,Agrawal2018} modelling language.}
    \label{fig:tradeoff_curves}
\end{figure}

 One can solve the SDP~\eqref{eq:budgetingproblem} for different values of $\gamma_{\textnormal{budget}}$ to obtain a relation between the diamond norm error and the $\gamma$-factor. This function, which we call \emph{tradeoff curve}, encapsulates the tradeoff between the approximation error and the sampling overhead. Example tradeoff curves for three gates are computed using a specific noise model and depicted in~\Cref{fig:tradeoff_curves}.
As expected, if the $\gamma$-factor budget is larger than the optimal $\gamma$-factor of the non-approximate QPD, the error becomes zero. Similarly, when the $\gamma$-factor budget is exactly $1$, then one does not gain any advantage over implementing the gate as-is without QEM. The more interesting regime is in between these two values of $\gamma_{\textnormal{budget}}$: One clearly sees that we can still significantly reduce the error of a gate, without having to pay the full $\gamma$-factor necessary for a non-approximate QPD. For the $\SWAP$ gate, the exact QPD requires a $\gamma$-factor of $2.21$ to completely correct the gate. However, if we only pay a $\gamma$-factor of $1.21$, we can still reduce the error by 67\%. The saved costs in terms of the sampling overhead are substantial, since the number of shots scales as $\gamma^{2 |C|}$, where $|C|$ is the number of gates.

If one applies the approximate quasiprobability method to a circuit with multiple gates, a new degree of freedom emerges, which is not present in the original formulation of the quasiprobability method: How much $\gamma$-factor budget do we give to every individual gate? Assume we have a budget $\gamma_{\textnormal{total}}$ for the whole circuit, how do we distribute that budget optimally across the whole circuit? More concretely, given $N$ gates we have to find individual budgets $\gamma_{\textnormal{budget},i}\geq 0$ for the $i$-th gate where $i=1,\dots,N$ such that \smash{$\prod_{i=1}^N{\gamma_{\textnormal{budget},i}} = \gamma_{\textnormal{total}}$}.
This is discussed in more detail in Appendix~\ref{app_resource_dist}.

\section{Stinespring algorithm}\label{sec_stinespring}
For a fixed decomposition set an optimal QPD can be found by solving the LP (\ref{eq:temme_lp}). To further reduce the $\gamma$-factor it would be advantageous to also find the optimal decomposition set for a given hardware noise. This task is non-trivial for at least two reasons:
\begin{enumerate}
    \item It is difficult to optimize over the space of all possible operations that a given quantum hardware can perform. In general this space is large and might vary significantly from one machine to another.\footnote{The space of all possible operations does not just consist of individual gates. It also comprises operations that use multiple gates, introduce ancilla qubits, perform measurements, trace out a subset of the qubits, etc. Basically any quantum circuit that starts and ends with the correct amount of qubits can be seen as a possible operation to be used in the decomposition set. Furthermore this space also depends on the exact noise that is present on the hardware.}
  \item Performing tomography to characterize the operations $\mathcal{E}_i$ is expensive. Hence, we want to limit the number of required uses of tomography.
\end{enumerate}

In this section, we will introduce a novel method, called the \emph{Stinespring algorithm}, that is able to deal with both issues. While the Stinespring algorithm is generally not able to produce the optimal decomposition set, it can be seen as an empirical approximation thereof. We demonstrate with simulations that the obtained decomposition set exhibits a significantly reduced $\gamma$-factor compared to the standard basis. The algorithm relies on a fundamental result in quantum information theory, stating that any quantum channel can be expressed as a unitary evolution on some extended Hilbert space.
\begin{theorem}[Stinespring dilation] \label{thm_stinespring}
  Consider a trace-preserving completelty positive (TPCP) map $\mathcal{E}\in \TPCP(A,A)$. There exists a Hilbert space $R$ and an isometry $V\in\linops(A,A\otimes R)$ such that for all density matrices $\rho$
  \begin{equation*}
      \mathcal{E}(\rho) = \tr_R[V\rho V^{\dagger}] \, .
  \end{equation*}
  Furthermore, there exists an isometry with $dim(R)\leq r$, where $r$ is the rank of the quantum channel defined by $r\vcentcolon = rank(\Lambda_{\mathcal{E}})$ for $\Lambda_{\mathcal{E}}$ the Choi matrix of $\mathcal{E}$.
\end{theorem}
Any isometry $V$ can be extended (generally non-uniquely) to a unitary $U\in\unitaryops(A\otimes R,A\otimes R)$ such that $U$ acts equivalently to $V$ on the space of states of the form $\rho_A\otimes \proj{0}_R$.

The core idea of the Stinespring algorithm is to find the set of general quantum channels $\{\mathcal{E}_i\}$ that allows the realization of a QPD with lowest possible $\gamma$-factors. These quantum channels are then approximately implemented on the quantum hardware by use of a Stinespring dilation.\footnote{We can straightforwardly use the Stinespring dilation to approximate an arbitrary quantum channel on a quantum computer by making use of clean ancilla qubits, performing a circuit corresponding to the unitary $U$ on the extended space, and finally discarding the ancilla qubits.} However, due to the significant hardware noise, the realized operation will only be an approximation of the desired quantum channel. The approximation error of the dilation is then accounted for in an iterative manner.

In practice, we will restrict the Stinespring algorithm to generate decomposition sets for error-mitigation of $1$-qubit and $2$-qubit gates, as the involved computation and tomography quickly become expensive for larger gates. This does not restrict the applicability of the method, as any multi-qubit unitary can be decomposed in single and two-qubit gates.

\subsection{Variational unitary approximation}
To optimize the approximate execution of a quantum channel via the Stinespring dilation, we make use of two central observations:
\begin{enumerate}
    \item Because the choice of the unitary $U$ extending the isometry $V$ is not unique we can try to choose the $U$ which requires the least amount of two-qubit gates to implement.
    \item Since the hardware is noisy, it may not be advantageous to implement a circuit that represents $U$ exactly. It might be better to use a circuit that approximates $U$ and requires less two-qubit gates and thus suffers less from noise on the hardware.
\end{enumerate}
In this section, we propose a method that makes use of both of these observations, which we call \emph{variational unitary approximation}. We use a variational circuit to implement the dilation unitary and fit its parameters in order to approximate the Stinespring isometry $V$ as well as possible. Denote by $\theta$ the tuple of all variational parameters and $U_{\mathrm{Var}}(\theta)$ the unitary represented by the variational form. Furthermore we denote by $V_{\mathrm{Var}}(\theta)$ the submatrix of $U_{\mathrm{Var}}(\theta)$ restricted on the subspace where the input ancilla qubits are fixed to the zero state. We want to choose our parameters $\theta$ such that we minimize the difference between $V_{\mathrm{Var}}(\theta)$ and $V$, where this optimization is done with a classical computer.
The variational unitary approximation allows us to freely choose the expressiveness of our variational circuit. More concretely, we can tune how many two-qubit gates it shall contain and therefore influence the tradeoff between the approximation error and the hardware noise error. Variational unitary approximations are studied in various settings such as~\cite{Khatri2019}, where the interested reader can find more information.

\begin{figure}[!htb]
\centering
    \tiny
    \quad 
    \Qcircuit @C=1em @R=.7em {
        \lstick{\ket{0}} & \gate{\mathrm{R_y}(\theta_1)} & \gate{\mathrm{R_z}(\theta_4)} & \ctrl{1} & \qw      & \gate{\mathrm{R_y}(\theta_7)} & \gate{\mathrm{R_z}(\theta_{10})} & \ctrl{1} & \qw      & \qw &        & & \ctrl{1} & \qw      & \gate{\mathrm{R_y}(\theta_{6m+1})} & \gate{\mathrm{R_z}(\theta_{6m+4})} & \qw \\
        \lstick{\ket{0}} & \gate{\mathrm{R_y}(\theta_2)} & \gate{\mathrm{R_z}(\theta_5)} & \targ    & \ctrl{1} & \gate{\mathrm{R_y}(\theta_8)} & \gate{\mathrm{R_z}(\theta_{11})} & \targ    & \ctrl{1} & \qw & \cdots & & \targ    & \ctrl{1} & \gate{\mathrm{R_y}(\theta_{6m+2})} & \gate{\mathrm{R_z}(\theta_{6m+5})} & \qw\\
        \lstick{\ket{0}} & \gate{\mathrm{R_y}(\theta_3)} & \gate{\mathrm{R_z}(\theta_6)} & \qw      & \targ    & \gate{\mathrm{R_y}(\theta_9)} & \gate{\mathrm{R_z}(\theta_{12})} & \qw      & \targ    & \qw &        & & \qw      & \targ    & \gate{\mathrm{R_y}(\theta_{6m+3})} & \gate{\mathrm{R_z}(\theta_{6m+6})} & \qw\\
        }
     \caption{Variational circuit used to approximate the the Stinespring unitary.}
     \label{fig_varForm}
\end{figure}
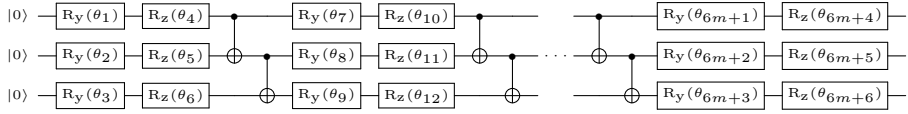

In the following we are going to look at a concrete example to illustrate the gains from the variational unitary approximation. Consider an arbitrary $2$-qubit quantum channel with rank $2$\footnote{The reason why we are interested in such a low-rank channel will be introduced in~\Cref{sec:rank_constrained_cd}.} that we want to approximate with a Stinespring dilation. Let $U$ be a $3$-qubit unitary realizing a Stinespring dilation of the given channel. The exact representation of $U$ typically requires around 100 $\CNOT$ gates (assuming linear qubit connectivity), which is significantly more than what current hardware can reasonably handle. We replace this ideal circuit with a $\mathrm{RyRz}$ variational circuit depicted in Figure~\ref{fig_varForm}, where $m$ denotes the depth of the variational form, the total number of parameters is $6(m+1)$ and the the total number of $\CNOT$ gates is $2m$. We use the gradient-based BFGS algorithm implemented in SciPy~\cite{Scipy} to minimize the objective $\norm{V - V_{\mathrm{Var}}(\theta)}_2$.\footnote{In order to avoid having to estimate the gradient with the finite differences method, we make use of the Autograd~\cite{Maclaurin2015} package for automatic differentiation.}  As initial guess for the parameters we use uniformly random numbers. The objective is non-convex and in practice the optimization algorithm finds a different local minimum depending on the provided initial guess. To obtain a good result in practice, we observed that it is enough to repeat the optimization 5 times for different initial values and then take best result.

\begin{figure}
    \centering
    \includegraphics[scale=1.0]{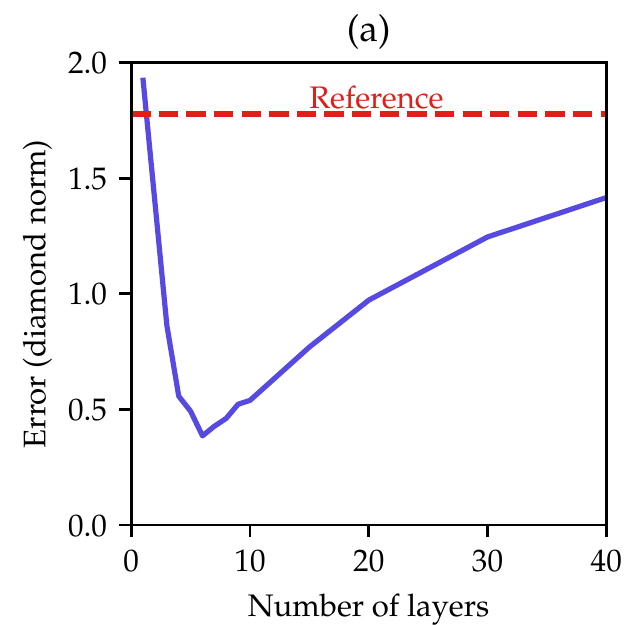} 
    \caption{Numerical results from the variational unitary approximation of a Haar-random $3$-qubit unitary $U$ using a $\mathrm{RyRz}$ variational form. This plot depicts the diamond norm error of the induced $2$-qubit quantum channel for different depths of the variational form. The depicted error contains the effects of the hardware noise and of the approximation error of the variational unitary. The noise model used is the same as in~\Cref{sec:tradeoff_curves}. The dashed red reference line corresponds to the error obtained when $U$ is (exactly) decomposed in single and two-qubit gates (without the variational unitary approximation) and then executed on the hardware.}
    \label{fig:variational_unitary_approximation}
\end{figure}

For purpose of numerical demonstration, the above procedure is performed for different depths of the variational form on a Haar-random $3$-qubit unitary $U$. By using the same noise model as in~\Cref{sec:tradeoff_curves}, we can estimate how well our method allows us to approximate the $2$-qubit quantum channel $\rho\mapsto \tr_3[U\rho U^{\dagger}]$ where $\tr_3$ stands for tracing out the third qubit. More precisely, we can compute the diamond norm error between the obtained $2$-qubit quantum channel and the ideal $2$-qubit quantum channel. This error encapsulates the approximation error of the variational form combined with the hardware noise. The results can be seen in~\Cref{fig:variational_unitary_approximation}. The variational unitary approximation technique allows us to significantly reduce the error to roughly one quarter of its reference value. One can also clearly see that there is a sweet spot in the variational circuit depth. This makes sense intuitively when considering the tradeoff mentioned previously: if the depth is too short, then $U$ is not well represented by the variational form and if the depth is too long, then the hardware noise takes over and we start to mostly sample noise. The optimum is reached at a variational circuit depth of $6$. 

\subsection{Channel decomposition}\label{sec:rank_constrained_cd}
Consider a trace\hyp{}preserving linear operator $\mathcal{F}\in \linops(\linops(A),\linops(B))$ of which we want to find an optimal QPD. In this section we will generalize the LP~(\ref{eq:temme_lp}) by not only optimizing over the quasiprobability coefficients $a_i$, but also over the decomposition set $\{\mathcal{E}_i\}_i$. The resulting QPD will not be directly usable for the quasiprobability method, as the obtained decomposition will generally not be implementable by a quantum computer in practice. To highlight this, we will denote the decomposition set obtained by this optimisation procedure by$\{\mathcal{G}_i\}_i$ instead of $\{\mathcal{E}_i\}_i$ throughout this section. We call the resulting QPD the channel decomposition of $\mathcal{F}$. We note that this decomposition was independently derived in~\cite{Piv_masterThesis,JWW20}. The channel decomposition will be an important step towards the realization of the Stinespring algorithm.

We denote the Choi matrix of $\mathcal{F}$ by ${\Lambda}_{\mathcal{F}}$ and our goal is now to construct a finite set of Choi matrices $\{\Lambda_{\mathcal{G}_i}\}_i$ which correspond to a decomposition set $\{\mathcal{G}_i\}_i$. The Choi representation is very convenient, as it allows us to formulate the TPCP-condition on the decomposition basis in a straightforward way: $\Lambda_{\mathcal{G}_i}\geq 0$ and $\tr_2[\Lambda_{\mathcal{G}_i}]=\frac{1}{2^n}\mathbb{1}$ for all $i$ where $\tr_2$ stands for the partial trace over the ancillary Hilbert space of the Choi-Jamiolkowski isomorphism and $n$ is the number of qubits involved in $\mathcal{F}$. The optimization problem at hand can be expressed as
\begin{align}\label{eq:cdd_firstattemt}
    \left \lbrace
    \begin{array}{r l}
        \min \limits_{a_i\in\mathbb{R},\Lambda_{\mathcal{G}_i}\in\mathbb{C}^{4^n\times 4^n}} & \sum\limits_{j=1}^{N} |a_j| \\
        \textnormal{s.t.} & \Lambda_\mathcal{F} = \sum\limits_{j=1}^N{a_j\Lambda_{\mathcal{G}_j}} \\
        &\Lambda_{\mathcal{G}_i}\geq 0 \\
        &\tr_2[\Lambda_{\mathcal{G}_i}]=\frac{1}{2^n}\mathbb{1} \, .
    \end{array} \right .
\end{align}
By grouping all positive and negative quasiprobability coefficients together
\begin{align*}
    a^+\coloneqq\sum_i{a_i^+}, 
    \quad a^-\coloneqq\sum_i{a_i^-} \, , 
    \quad \mathcal{G}^+\coloneqq\frac{1}{a^+}\sum_i{a_i^+\mathcal{G}_i^+}, 
    \quad \textnormal{and} \quad \mathcal{G}^-\coloneqq\frac{1}{a^-}\sum_i{a_i^-\mathcal{G}_i^-} \, ,
\end{align*}
one can see that choosing $N=2$ is enough to reach the optimum. By performing the substitution $\tilde{\Lambda}^{\pm}\coloneqq a^\pm \Lambda_{\mathcal{G}^\pm}$ we can transform the problem into a SDP which can be solved efficiently
\begin{align*}
    \left \lbrace
    \begin{array}{r l}
        \min \limits_{a_i\in\mathbb{R},\Lambda_{\mathcal{G}_i}\in\mathbb{C}^{4^n\times 4^n}} & a^+ + a^- \\
        \textnormal{s.t.} & \Lambda_\mathcal{F} = \tilde{\Lambda}_{\mathcal{G}^+} - \tilde{\Lambda}_{\mathcal{G}^-} \\
        &\tilde{\Lambda}_{\mathcal{G}^{\pm}}\geq 0 \\
        &\tr_2[\tilde{\Lambda}_{\mathcal{G}^{\pm}}]=a^{\pm}\frac{1}{2^n}\mathbb{1} \, .
    \end{array} \right .
\end{align*}
The channel decomposition asserts that we can find TPCP maps $\mathcal{G}^+,\mathcal{G}^-$ and $a^+,a^-\geq 0$ such that $\mathcal{F}=a^+\mathcal{G}^+ - a^-\mathcal{G}^-$ with optimal $\gamma$-factor $\gamma=a^++a^-$. 

For the sake of the Stinespring algorithm, we would like to approximate the $\mathcal{G}^{\pm}$ using a Stinespring dilation. This will only work reasonably well when the number of required ancilla qubits is not too large. We can enforce this by adding an additional constraint $\rank(\Lambda_{\mathcal{G}_i}^{\pm})\leq r$ for some $r\in\mathbb{N}$ into the SDP (\ref{eq:cdd_firstattemt}), i.e., 
\begin{align}\label{eq:rank_constrained_cd}
    \left \lbrace
    \begin{array}{r l}
        \min \limits_{a_i^{\pm}\in\mathbb{R}^+,\tilde{\Lambda}_{\mathcal{G}_i}^{\pm}\in\mathbb{C}^{4^n\times 4^n}} & \sum\limits_{i=1}^{n_{\mathrm{pos}}} a_i^+  +  \sum\limits_{i=1}^{n_{\mathrm{neg}}}{a_i^-} \\
        \textnormal{s.t.} & \Lambda_\mathcal{F} = \sum\limits_{i=1}^{n_{\mathrm{pos}}}{\tilde{\Lambda}_{\mathcal{G}_i}^+} - \sum\limits_{i=1}^{n_{\mathrm{neg}}}{\tilde{\Lambda}_{\mathcal{G}_i}^-} \\
        &\tilde{\Lambda}_{\mathcal{G}_i}^{\pm}\geq 0 \\
        &\tr_2[\tilde{\Lambda}_{\mathcal{G}_i}^{\pm}]=a_i^{\pm}\frac{1}{2^n}\mathds{1} \\
        &\rank(\tilde{\Lambda}_{\mathcal{G}_i}^{\pm})\leq r \, ,
    \end{array} \right .
\end{align}
where $\tilde{\Lambda}_{\mathcal{G}_i}^{\pm}\coloneqq a_i^\pm \Lambda_{\mathcal{G}_i}^\pm$. Note that with the rank constraint we do not have the guarantee anymore that we can decompose $\mathcal{F}$ into just two channels, so generally it is not evident how small we can choose the number of positive and negative channels, $n_{\mathrm{pos}}$ and $n_{\mathrm{neg}}$, respectively, while still finding the optimum. Because SDPs with rank constraints are NP-hard, we have to resort to heuristics to find a solution. One common approach is due to Burer-Monteiro~\cite{Burer2003,Burer2005}. Suppose that we want to solve a given SDP with a rank constraint $\rank(C)\leq r$ for some positive-semidefinite $n\times n$ matrix $C$. The main idea is to parametrize $C=X^{\dagger} X$ for some $r\times n$ complex matrix $X$ and then optimize over the matrix elements of $X$. The positive-semidefiniteness and rank constraint of $C$ are automatically enforced from the construction. Unfortunately, the objective and the constraints generally contain quadratic terms of $X$, so the problem becomes a quadratically-constrained quadratic program. Still, recent research has demonstrated that the objective landscape of these problem tends to behave nicely and that using local optimization methods can provably lead to the global optimum under some assumptions~\cite{Cifuentes2019,Boumal2016}. We defer further technical details to~\Cref{app_solver}.

\subsection{The algorithm}\label{sec:stinespring_algo}
The result in \Cref{sec:rank_constrained_cd} allows us to quasiprobabilistically decompose any TPCP map $\mathcal{F}$, which could for example be our desired unitary $\mathcal{U}$, into rank $r$ quantum channels that can be approximated with $\ceil{\log_2{r}}$ ancilla qubits. By choosing $r$ small enough and by making use of the variational unitary approximation, we can ensure that this approximation is reasonably good.
Still, there is an important step missing before we can practically use this result. The quasiprobability method requires a QPD where the elements of the decomposition set correspond to channels describing noisy operations which the actual quantum hardware can execute. However, due to the noisy nature of the hardware, we can only execute an approximation of the channels $\mathcal{G}_i$ obtained from the channel decomposition. We have to take into account the inaccuracy of the implemented Stinespring dilation when constructing our decomposition set. This will be achieved by the use of an iterative algorithm.

Assume that we have access to a noise oracle $\mathcal{G}\mapsto\mathcal{N}(\mathcal{G})$ which gives us the actual quantum channel executed by the hardware when we implement the channel $\mathcal{G}$ using a Stinespring dilation. In general, since we are considering single and two-qubit gates, this oracle can be realized by simply performing tomography of the noisy realisations of the $\mathcal{G}_i$ on the hardware. Assume that we found a channel decomposition
\begin{equation*}
    \mathcal{F} = \sum_{i=1}^{n_{\mathrm{pos}}}{a_i^+\mathcal{G}_i^+} - \sum_{i=1}^{n_{\mathrm{neg}}}{a_i^-\mathcal{G}_i^-}
\end{equation*}
using the rank-constrained optimization~\eqref{eq:rank_constrained_cd}. If we were to implement $\mathcal{F}$ with the quasiprobability method by using the Stinespring approximation of the involved channels $\mathcal{G}_i^{\pm}$, an error $\delta$ would occur where
\begin{equation*}
    \delta \vcentcolon = \mathcal{F} - \sum_{i=1}^{n_{\mathrm{pos}}}{a_i^+\mathcal{N}(\mathcal{G}_i^+)} - \sum_{i=1}^{n_{\mathrm{neg}}}{a_i^-\mathcal{N}(\mathcal{G}_i^-)} \, .
\end{equation*}
We can iteratively repeat our procedure, but this time decomposing $\delta$ instead of $\mathcal{F}$. During each point of the iteration we store the $\mathcal{N}(\mathcal{G}_i^{\pm})$ into our decomposition set. The $\delta$ can be obtained by performing an approximate QPD of the target unitary channel $\mathcal{U}$ into the decomposition set. At some point our decomposition set will be large enough such that the resulting approximation error is smaller than a desired threshold. A conceptual visualization of the procedure is depicted in~\Cref{fig:stinespring_iteration} and the exact algorithm can be found in Algorithm~\ref{algo:stinespring}.

\begin{figure}[!htb]
    \centering
\begin{tikzpicture}
\draw[thick,->] (0,0) -- (1,3);
\node at (0.25,2) {$\cF$};

\draw[thick,->,color={rgb:red,41;green,191;blue,18}] (0,0) -- (1.5,1.5);
\draw[thick,->,color={rgb:red,41;green,191;blue,18}] (1.5,1.5) -- (1,3);
\node[color={rgb:red,41;green,191;blue,18},rotate=45] at (0.75,1.1) {$a_1 \cE_1$};
\node[color={rgb:red,41;green,191;blue,18},rotate=282] at (1.1,2) {$a_2 \cE_2$};

\draw[thick,->,color={rgb:red,2;green,165;blue,207}] (0,0) -- (1.8,1.1);
\draw[thick,->,color={rgb:red,2;green,165;blue,207}] (1.8,1.1) -- (1.5,2.8);
\node[color={rgb:red,2;green,165;blue,207},rotate=30] at (1.2,0.4) {$a_1 \cN(\cE_1)$};
\node[color={rgb:red,2;green,165;blue,207},rotate=282] at (1.9,2) {$a_2 \cN(\cE_2)$};

\draw[thick,->,color={rgb:red,222;green,32;blue,25}] (1.5,2.8) -- (1,3);
\node[color={rgb:red,222;green,32;blue,25}] at (1.3,3.1) {$\delta$};

\node at (7.55,3) {\textbf{{\color[RGB]{41,191,18}Step 1:}} Compute channel decomposition $\cF = \sum_i a_i \mathcal{G}_i$};
\node at (7.65,2.5) {into quantum channels $\mathcal{G}_i$ of rank $\leq r$};

\node at (7.85,1.9) {\textbf{{\color[RGB]{2,165,207}Step 2:}} Realize $\mathcal{G}_i$ using Stinespring dilation and use noise};
\node at (8.22,1.4) {oracle to find error caused by hardware noise};

\node at (5.35,0.8) {\textbf{{\color[RGB]{222,32,25}Step 3:}} Determine error $\delta$};

\node at (7.23,0.2) {\textbf{Step 4:} Jump to Step 1 with $\delta$ as target operation};

\end{tikzpicture}
    \caption{Graphical visualization of the iterative process involved in the Stinespring algorithm.}
    \label{fig:stinespring_iteration}
\end{figure}

\begin{algorithm}[ht]\label{algo:stinespring}
    \SetAlgoLined
    \textbf{Given:} a target unitary operation $[U]$ for $U\in\unitaryops(A)$, a noise oracle $\mathcal{N}$, a threshold $\Delta_{\textnormal{threshold}}$ for the diamond norm error\;
    \KwResult{A decomposition set for a QPD with low sampling overhead}
    $\mathcal{D} \vcentcolon = \{\mathcal{N}([U])\}$\;
    $\mathcal{F} \vcentcolon = [U]$\;
    \Repeat {
        $\{(a_j,D_j)\}_j \vcentcolon = $ compute approximate QPD coefficients of $\mathcal{F}$ using decomposition set $\mathcal{D}$\;
        $\Delta \vcentcolon = $ get diamond norm error of approximate QPD $\{(a_j,D_j)\}_j$\;
        \If{$\Delta<\Delta_{\textnormal{threshold}}$}{
            break\;
        }
        $\delta \vcentcolon = $ get the remaining error of the approximate QPD $\{(a_j,D_j)\}_j$\;
        $\{\mathcal{G}_i\}_i \vcentcolon = $ perform rank-constrained channel decomposition of $\delta$ into a set of channels\;
        $\{V_i\}_i \vcentcolon = $ get Stinespring dilation isometries of the channels $\{\mathcal{G}_i\}_i$\;
        $\{C_i\}_i \vcentcolon = $ get variational unitary approximation circuits of the isometries $\{V_i\}_i$\;
        $\{\mathcal{E}_i\}_i \vcentcolon = $ apply noise oracle $\mathcal{N}$ on the circuits $\{C_i\}_i$\;
        $\mathcal{D} \vcentcolon = \mathcal{D} \cup \{\mathcal{E}_i\}_i$\;
    }
    \Return $\mathcal{D}$
    \caption{Stinespring algorithm}
\end{algorithm}

It is crucial that the Stinespring dilation allows for a reasonably accurate approximation of a desired quantum channel. In other words, the blue and green arrows in~\Cref{fig:stinespring_iteration} have to be sufficiently close. In the simulations presented in \Cref{sec_simulationResults}, we observed that this was only the case with the inclusion of the rank constraint in the channel decomposition. The variational unitary approximation technique significantly improves the approximation further.


\subsection{Simulation results} \label{sec_simulationResults}
\begin{figure}[h]
    \centering
    \includegraphics{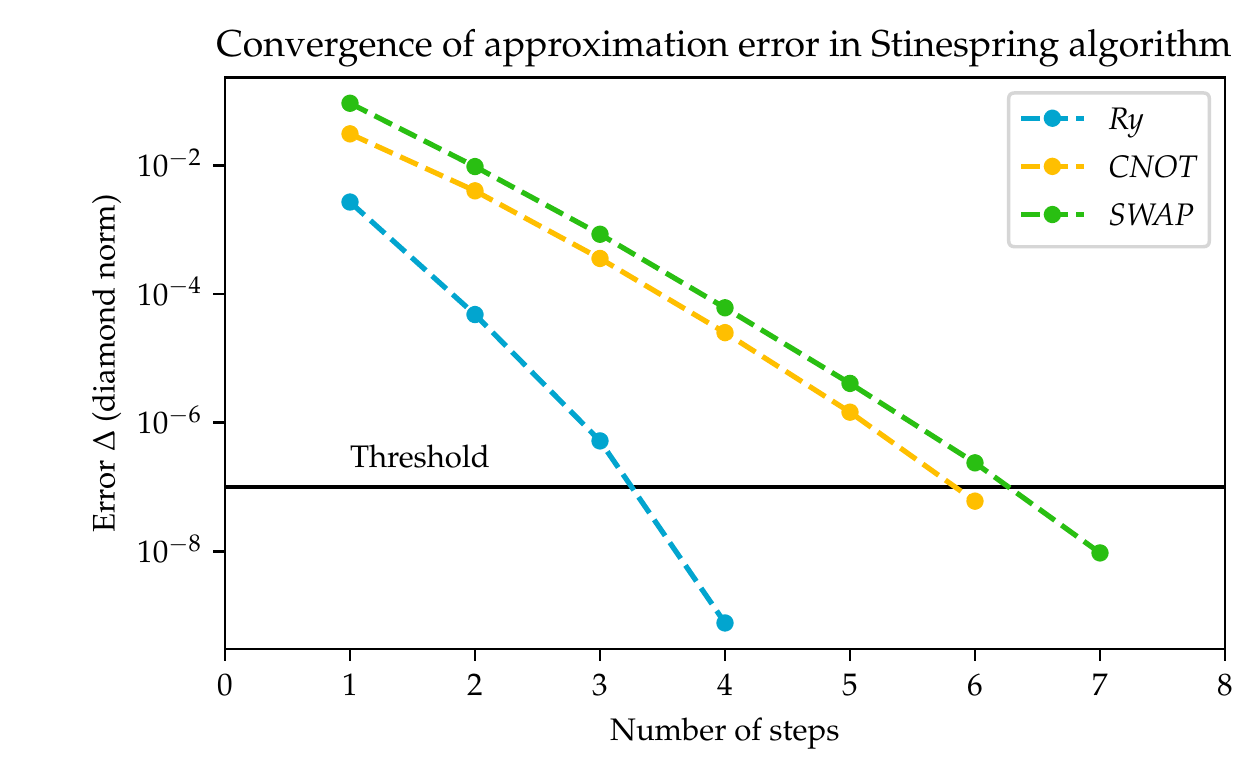}
    \caption{Evolution of the approximation error in each step of the Stinespring algorithm. Three different runs of the algorithm on three different gates $\Ry$, $\CNOT$ and $\SWAP$ are depicted. The hardware noise is estimated using a noise model approximating the \texttt{ibmq\_melbourne} device. The horizontal grey line denotes the threshold of $10^{-7}$ at which the algorithm stops.}
    \label{fig:stinespring_convergence}
\end{figure}

We next show the performance of the Stinespring algorithm via simulation results on the gates $\Ry$, $\CNOT$, and $\SWAP$, that we already analyzed in~\Cref{sec:tradeoff_curves}. For all three gates we enforce a rank-constraint $r=2$ during the channel decomposition. This corresponds to allowing for at most one ancilla qubit during the Stinespring dilation. For the two-qubit gates we additionally make use of the variational unitary approximation with a $3$-qubit $\mathrm{RyRz}$ variational form of depth 6 with linear qubit connectivity. The noise oracle is obtained from the noise model that was already used in~\Cref{sec:tradeoff_curves}, which aims to approximate the noise on the \texttt{ibmq\_melbourne} device. Using a noise model instead of a full tomography significantly speeds up the simulation. We use a threshold of $\Delta_{\textnormal{threshold}}=10^{-7}$ to stop the iterative procedure.
During each iteration of the Stinespring algorithm, we store the diamond norm error of the current approximate QPD (denoted $\Delta$ in Algorithm~\ref{algo:stinespring}). \Cref{fig:stinespring_convergence} shows how this error decreases during the Stinespring algorithm. It is clearly visible that this decrease is exponential.

As discussed in detail in~\Cref{app_solver}, each iteration of the Stinespring algorithm applied on the 2-qubit gates extends the decomposition set by 16 new operations. By considering that we need around 6-7 steps to reach the desired threshold, this implies that the produced decomposition set is significantly smaller than the standard basis (around 70-80 elements instead of 256). This result is remarkable and indicates that the Stinespring algorithm really does find a decomposition set that is well adapted to the hardware noise. As a reminder, the 256 elements in the standard basis were needed to completely span the space of Hermitian-preserving operators. The decomposition set produced by the Stinespring algorithm spans a significantly smaller space, as it is tailored to only represent one specific operation.

~\Cref{tab:stinespring_result} denotes the $\gamma$-factors obtained by finding the optimal QPD using the decomposition set produced by the Stinespring algorithm for the three gates in question. We see that a significant improvement can be observed compared to using the decomposition set consisting of the standard basis and the respective noisy gate. As previously discussed, the sampling overhead of the quasiprobability method scales exponentially in the number of gates, where the $\gamma$-factor forms the basis of that exponential cost. Therefore any reduction in the $\gamma$-factor is significant and allows us to implement considerably larger quantum circuits.

Consider for instance that we have a circuit consisting of gates with equal sampling overhead $\gamma=1+\kappa$ and we wish to fix the total sampling overhead $\Gamma=(1+\kappa)^{2n_{\textnormal{gates}}}$. Then the number of gates $n_{\textnormal{gates}}$ that one can error mitigate without exceeding the total sampling overhead $\Gamma$ is given by \smash{$\frac{1}{2}\frac{\ln{\Gamma}}{\ln{1+\kappa}}$} which is \smash{$\approx \frac{1}{2}\frac{\ln{\Gamma}}{\kappa}$} for $\kappa\approx 0$. So a reduction of $\kappa$ by a factor $2$, which is roughly what we achieved for the $\Ry$ and $\CNOT$ gates, corresponds to doubling the number of gates that we can error mitigate.

Finally, we repeat the experiment on the \emph{CNOT} gate for two additional noise models. We again consider noise models from Qiskit which approximate the noise on the \texttt{ibmq\_sydney} and \texttt{ibmq\_mumbai} hardware platforms. These two chips are newer and more advanced than \texttt{ibmq\_melbourne} and therefore exhibit lower noise. The results are depicted in~\cref{tab:other_noisemodels} and one sees that the Stinespring algorithm again provides a decomposition set with a significantly reduced $\gamma$-factor.

We do not have a guarantee that the Stinespring algorithm always converges. Such a proof would require a formalization of the heuristics that were made during the construction of the algorithm. In Appendix~\ref{app_heuristicts} we summarize these heuristics in detail.

\begin{table}[ht]
    \centering
    \begin{tabular}{|c|c|c|}
        \hline
        & \multicolumn{2}{c|}{\phantom{.}$\!\!\!\!\!\!\!\!\!\!\!\!\!\!\!\!\!\!\!\gamma$-factor} \\
        & standard basis + noisy gate & Stinespring algorithm \\
        \hline
        $\Ry$ & 1.0106 & 1.0056 \\
        $\CNOT$ & 1.1789 & 1.0815 \\
        $\SWAP$ & 2.2095 & 1.2294 \\
        \hline
    \end{tabular}
    \caption{Simulation results of the Stinespring algorithm applied on the three quantum gates $\Ry$, $\CNOT$, and $\SWAP$. The obtained $\gamma$-factor is compared to the value obtained when the respective gates are decomposed into the standard basis.}
    \label{tab:stinespring_result}
\end{table}
\begin{table}[ht]
    \centering
    \begin{tabular}{|l|l|c|c|}
        \hline
        hardware & $\CNOT$ & \multicolumn{2}{c|}{\phantom{.}$\!\!\!\!\!\!\!\!\!\!\!\!\!\!\!\!\!\!\!\gamma$-factor} \\
        backend & error rate & standard basis + noisy CNOT & Stinespring algorithm \\
        \hline
        \texttt{ibmq\_melbourne} & 2.03\% & 1.1789 & 1.0815 \\
        \texttt{ibmq\_mumbai} & 1.12\% & 1.0968 & 1.0486 \\
        \texttt{ibmq\_sydney} & 0.96\% & 1.0667 & 1.0415 \\
        \hline
    \end{tabular}
    \caption{Simulation results of the Stinespring algorithm applied on the $\CNOT$ gate using noise models from different IBMQ hardware backends. The obtained $\gamma$-factor is compared to the value obtained when the respective gates are decomposed into the standard basis.}
    \label{tab:other_noisemodels}
\end{table}

\section{Discussion}
The quasiprobability method is an error mitigation technique that allows us to suppress errors on small-scale devices without the need of implementing universal fault-tolerant quantum computing. The cost is a sampling overhead scaling as $\mathcal{O}(\gamma^{2|C|})$, where $\gamma\geq 1$ is the $\gamma$-factor and $|C|$ denotes the number of gates in the circuit that need to be error mitigated.  
We presented two novel methods to reduce the $\gamma$-factor. Since the sampling overhead scales exponentially in the number of gates with the $\gamma$-factor being the basis of the exponent, reducing $\gamma$ is crucial.

In this work, we presented two improvements of the QPD method that both allow to reduce the overhead by minimizing the $\gamma$-parameter. The first contribution, i.e., the approximate QPD, allows us to elegantly generalize the quasiprobability method in an approximate setting, where one can make a tradeoff between an approximation error and the sampling overhead. Since in many practical applications it is very difficult to obtain exact estimates of the hardware noise, the quasiprobability method will anyways incur a significant error. In this setting it might be possible to reduce the sampling overhead without significantly increasing the error by making use of our approximate quasiprobability method.

Whereas the approximate QPD approach is fully analytical and mathematically rigorous, the second improvement (the Stinespring algorithm) relies on some numerical heuristics that we tested on various practical problems and verified its usefulness via simulation results. Nevertheless it would be important to prove the convergence or even the convergence rate of the Stinespring algorithm. We leave this for future work. Our results indicate that this approach could have the potential to significantly reduce the sampling overhead compared to existing methods, as can be seen from a simple calculation. Suppose we have a quantum circuit where we want to correct as many $\SWAP$ gates as possible with a total sampling overhead that is not larger than $10^3$.\footnote{Note that due to limited connectivity on near-term devices $\SWAP$ gates will occur very often.} Table~\ref{tab:stinespring_result} shows that via the previous QPD approach we can correct $4$ $\SWAP$ gates, whereas our improved QPD method allows us to correct up to $16$ $\SWAP$ gates.\footnote{Recall that the overhead scales as $\mathcal{O}(\gamma^{2 |C|})$, where $|C|$ is the number of mitigated gates.} From an experimental point of view this technique also has the advantage that, in contrary to the standard basis, no measurements have to be executed on the data qubits.

Due to its inherent exponential scaling, neither the QPD nor other error mitigation techniques are believed to be able to replace the need for error correction for complicated quantum computations. However error mitigation techniques may be able to smoothen the transition into the fault-tolerant quantum computing era. Hence as a next step for future work it would be interesting to see how the improved QPD techniques presented in this work can be combined with error correction ideas.

\paragraph{Author contributions} All authors contributed equally to this work.

\paragraph{Competing interests} The authors declare that there are no competing interests.
\paragraph{Acknowledgements}
We thank Sergey Bravyi, Jay Gambetta, and Kristan Temme for discussions on quasiprobability decompositions. 
IBM, the IBM logo, and ibm.com are trademarks of International Business Machines Corp., registered in many jurisdictions worldwide. Other product and service names might be trademarks of IBM or
other companies. The current list of IBM trademarks is available at \url{https://www.ibm.com/legal/copytrade}.
\paragraph{Code availability}
The code to redproduce all numerical experiments and plots in this paper is available at \url{https://github.com/ChriPiv/stinespring-algo-paper}.
\appendix
\section{Optimal resource distribution} \label{app_resource_dist}
\Cref{sec_approx_QPD} motivates the question of how to distribute a given $\gamma$-factor budget on different individual gates.
Before we continue, we have to clarify what objective function we seek to optimize when we talk about an optimal budget distribution. The overall goal is to minimize the error (in terms of the diamond norm) of the total circuit. Denote by $\mathcal{F}_{\textnormal{approx},i}$ the channel produced by the approximate QPD of the $i$-th quantum gate and $U_i$ the unitary corresponding to the $i$-th gate, where $\cU_i(\cdot)= U_i (\cdot) U^\dagger_i$. 
The objective function is therefore
\begin{equation}\label{eq:totalerror}
    \norm{\mathcal{F}_{\textnormal{approx},N}\circ\dots\circ\mathcal{F}_{\textnormal{approx},2}\circ\mathcal{F}_{\textnormal{approx},1} - \cU_N\circ\dots\circ \cU_2\circ \cU_1}_{\diamond} \, .
\end{equation}
The computation of this quantity is intractable, as it would require to simulate the complete noisy circuit. Therefore, we need to find a proxy that is computable from local quantities.

From the triangle inequality and the fact that the diamond norm is a contraction under completely positive maps, one immediately obtains
\begin{align*}
\norm{\mathcal{E}\circ\mathcal{F}-\mathcal{E'}\circ\mathcal{F'}}_{\diamond}
\leq \norm{\mathcal{E}\circ\mathcal{F}-\mathcal{E'}\circ\mathcal{F}}_{\diamond} + \norm{\mathcal{E'}\circ\mathcal{F}-\mathcal{E'}\circ\mathcal{F'}}_{\diamond} 
\leq \norm{\mathcal{E}-\mathcal{E'}}_{\diamond} + \norm{\mathcal{F}-\mathcal{F'}}_{\diamond} \, ,
\end{align*}
for any completely positive maps $\mathcal{E},\mathcal{F},\mathcal{E}',\mathcal{F}'$ where $\mathcal{E},\mathcal{E}'$ are trace-preserving. If we wish to apply this bound to our setting in~\Cref{eq:totalerror}, we have to guarantee that the $\mathcal{F}_{\textnormal{approx},i}$ are TPCP. This can be achieved by inserting an additional constraint into the SDP of the approximate QPD. Using the notation $\varepsilon_i(\cdot)$ to denote the tradeoff curve of the $i$-th gate, we thus obtain the optimization problem
\begin{align}\label{eq:global_budgeting}
    \left \lbrace
    \begin{array}{r l}
        \min \limits_{\gamma_{\textnormal{budget},i}\in\mathbb{R}} & \sum_j{\varepsilon_j(\gamma_{\textnormal{budget},j})} \\
        \textnormal{s.t.}  & \gamma_{\textnormal{budget},i}\geq 0 \, \forall i \\ 
        & \prod_j {\gamma_{\textnormal{budget},j}} = \gamma_{\textnormal{total}}\, .
    \end{array} \right .
\end{align}
Problem~\eqref{eq:global_budgeting} is generally non-convex, as we have not enough guarantees on the shape of the tradeoff curves. Still in practice we observe that the objective is convex in a broad region around the optimum, so we expect numerical heuristics based on gradient descent to work well.

\begin{figure}
    \centering
    \includegraphics{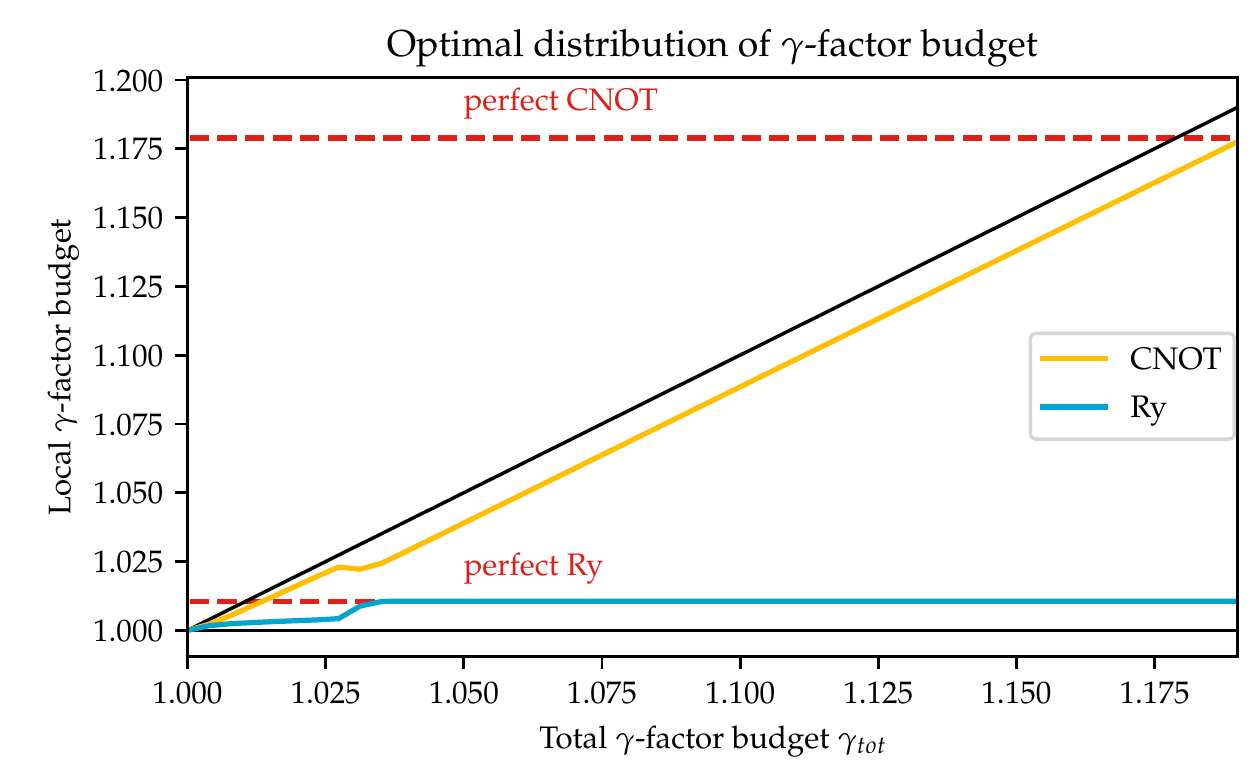}
    \caption{This curve shows the optimal $\gamma$-factor budget distribution across a circuit consisting of a single $\Ry$ and $\CNOT$ gate. The total $\gamma$-factor budget $\gamma_{\mathrm{total}}$ is varied between $1$ and $1.191$, which is the minimal value necessary to obtain perfect QPD for both gates at the same time. The red dashed lines denote the necessary $\gamma$-factor to decompose the respective gate perfectly. The black line denotes $\gamma_{\mathrm{total}}$ and therefore corresponds to the multiplication of the blue and yellow curves.}
    \label{fig:global_budgeting}
\end{figure}

We finish this section with a small demonstration on a simple circuit that consists of a $\Ry$ and a $\CNOT$ gate. The tradeoff curves are computed exactly as in~\Cref{sec:tradeoff_curves}, with the sole difference that we now include a TPCP-constraint for the approximate QPD. We evaluate the tradeoff curves at discrete points and use linear interpolation in order to extrapolate this data to a full function that can be used in the optimization routine. We solve the optimization problem~(\ref{eq:global_budgeting}) for different values of $\gamma_{\textnormal{total}}$ using a black box solver based on the BFGS algorithm implemented in the SciPy~\cite{Scipy} software package. The results are depicted in~\Cref{fig:global_budgeting}. Interestingly, the optimal strategy for budget distribution is nontrivial. In the regime with a small budget $1\leq \gamma_{\textnormal{total}}\lessapprox 1.02$, it is optimal to give most of the $\gamma$-factor budget solely to the $\CNOT$ gate. In a transition regime $1.02\lessapprox \gamma_{\textnormal{total}}\lessapprox 1.03$ both gates obtain a significant amount of $\gamma$-factor budget. In the upper regime $1.03\lessapprox \gamma_{\textnormal{total}}$ the $\Ry$ gate is perfectly decomposed and the remaining budget is then given to the $\CNOT$ gate. So whether one should prioritize the $\CNOT$ or the $\Ry$ gate in the budgeting strongly depends on the value of $\gamma_{\textnormal{total}}$.
\section{Details on rank-constrained channel decomposition}\label{app_solver}
In our case the $\Lambda_{\mathcal{G}_i}^{\pm}$ do not show up in the objective function, so we actually have a quadratically-constrained linear program, i.e.~a problem of the form
\begin{align}\label{eq:bm_setting}
    \min \limits_{x\in \R^n} \{ f(x):  g(x)=0 \} \, ,
\end{align}
where we grouped all variables into a large vector $x$, $f$ is a linear function and $g$ is a quadratic map that encapsulates all constraints. Furthermore, we know the optimum $f^*$ of the objective function without the rank constraint. Let's make an assumption for the moment being:
\begin{assumption}\label{ass:bm_rank_solution}
  The rank-constrained optimization problem~\eqref{eq:rank_constrained_cd} reaches the same minimal $\gamma$-factor as the original SDP without the rank constraint.
\end{assumption}
This assumption does obviously not hold for $r=1$, as it would imply that we could decompose any $\mathcal{F}\in \TP(A,A)$ into unitary operations. We are only concerned with the case $r\geq 2$ in practice, so let us assume for the moment that the assumption is correct. This allows us to reformulate~\Cref{eq:bm_setting} as
\begin{align}\label{eq:bm_setting2}
        \min \limits_{x \in \R^n} \{ \norm{g(x)}_2^2: f(x)=f^* \} \, .
\end{align}
We say that our optimization only succeeds if it finds a solution of problem~\eqref{eq:bm_setting2} which minimizes the quadric objective $\norm{g(x)}_2^2$ to zero. The reason why we swap objective and constraint is because we observed linear constraints to be easier to deal with numerically than quadratic (and possibly nonconvex) constraints.

To numerically solve problem~(\ref{eq:bm_setting2}) we made use of the trust-region algorithm for constrained optimization which is included in SciPy~\cite{Scipy}. We fixed $r=2$, which corresponds to allowing at most one ancilla qubit in the Stinespring dilation. In the single-qubit case\footnote{Here we refer to the number of qubits $n$ on which $\mathcal{U}$ acts.} we could consistently find a good solution, whereas in the two-qubit case we were often confronted with convergence problems. It is not exactly clear whether these problem stemmed from the non-existence of a low-rank solution (i.e.~\Cref{ass:bm_rank_solution} being wrong) or from numerical issues with the Burer-Monteiro method. In any case, we observed that the convergence could be significantly improved by allowing some error in the linear constraint, i.e.,
\begin{align*}
\min \limits_{x \in \R^n} \{ \norm{g(x)}_2^2 :  f(x)\leq(1+\varepsilon) f^*  \} \, ,
\end{align*}
where we used a value of $\varepsilon =0.2$. In other words, we allow for a sub-optimality in the $\gamma$-factor by at most 20\%. In principle one could try to find the smallest admissible $\epsilon$ by usage of bisection. We used $n_{\mathrm{pos}}=n_{\mathrm{neg}}=2$ for single-qubit quantum channels and $n_{\mathrm{pos}}=n_{\mathrm{neg}}=8$ for two-qubit quantum channels.

We also observed a good initialization for problem~(\ref{eq:bm_setting2}) to be of major importance. In the rest of the section we focus on this aspect. The task at hand is to solve following optimization problem
\begin{align*}
    \left \lbrace
    \begin{array}{r l}
        \min \limits_{a_i^{\pm}\in\mathbb{R}^+,X_i^{\pm}\in\mathbb{C}^{2\times 4^n}} & \norm{\Lambda_\mathcal{F} \!-\! (\sum\limits_{j=1}^{n_{\mathrm{pos}}}{\tilde{\Lambda}_{\mathcal{G}_j}^+} \!\!-\!\! \sum\limits_{j=1}^{n_{\mathrm{neg}}}{\tilde{\Lambda}_{\mathcal{G}_j}^-})}_2^2 \!\!+\!\!\sum\limits_{j=1}^{n_{\mathrm{pos}}}\norm{\tr_2[\tilde{\Lambda}_{\mathcal{G}_j}^{+}]\!-\!a_j^+\frac{1}{2^n}\mathds{1}}_2^2 \!+\! \sum\limits_{j=1}^{n_{\mathrm{neg}}}\norm{\tr_2[\tilde{\Lambda}_{\mathcal{G}_j}^{-}]\!-\!a_j^-\frac{1}{2^n}\mathds{1}}_2^2 \\
        \textnormal{s.t.} & f^* \leq \sum_{j=1}^{n_{\mathrm{pos}}} a_j^+ + \sum_{j=1}^{n_{\mathrm{neg}}} a_j^- \leq f^*(1+\epsilon) \\
        &\tilde{\Lambda}_{\mathcal{G}_i}^{\pm} = (X_i^{\pm})^{\dagger} X_i^{\pm} \, .
    \end{array} \right .
\end{align*}
We restrict ourselves to the case of $r=2$, which corresponds to allowing a single ancilla qubit in the Stinespring dilation. In order to solve this problem with a local method, an initial guess $a_i^{\pm,0},X_i^{\pm,0}$ for the parameters is required. We present a heuristic to find such initial values which seems to work well in our experience.
We start off by computing the channel difference decomposition $\mathcal{F}=a^+\mathcal{G}^+ - a^-\mathcal{G}^-$ of our target operation. We consider the spectral decomposition of the Choi matrices of $\mathcal{G}^{\pm}$
\begin{align}
    \Lambda_{\mathcal{G}^{\pm}} = \begin{pmatrix} \vert & & \vert \\ u_1^{\pm} & \dots & u_{4^n}^{\pm} \\ \vert & & \vert \end{pmatrix}
     \mathrm{diag}(\lambda_1,\dots,\lambda_{4^n})
    \begin{pmatrix} \text{---} & (u_1^{\pm})^{\dagger} & \text{---} \\ & \vdots & \\ \text{---} & (u_{4^n}^{\pm})^{\dagger} & \text{---} \end{pmatrix}
= \sum_{i=1}^{4^n}{ \lambda_i u_i^{\pm} (u_i^{\pm})^{\dagger} } \, , \label{eq:choimat_spectral_decomp}
\end{align}
where $\lambda_i^{\pm}$ denote the eigenvalues and $u_i^{\pm}$ denote the corresponding eigenvectors. We chose the indices such that the $\lambda_i^{\pm}$ are ordered non-increasingly in $i$. Equation~\Cref{eq:choimat_spectral_decomp} is a decomposition of $\mathcal{G}^{\pm}$ into rank-1 operations. Since we want to find a decomposition into rank-2 operations, we group these rank-1 matrices into pairs
\begin{align*}
    \Lambda_{\mathcal{G}^{\pm}} = \sum_{i=1}^{4^n/2}{ Y_i^{\pm} (Y_i^{\pm})^{\dagger} } \quad \textnormal{where} \quad
    Y_i = \begin{pmatrix} \vert & \vert \\ \sqrt{\lambda_{2i}}u_{2i}^{\pm} & \sqrt{\lambda_{2i+1}}u_{2i+1}^{\pm} \\ \vert & \vert \end{pmatrix} \, .
\end{align*}
We choose our initial guess as $a_i^{\pm,0} \vcentcolon = \tr[Y_i]$  and  $X_i^{\pm,0} \vcentcolon = \frac{Y_i}{\tr[Y_i]}$.
If $n_{\mathrm{pos}}=n_{\mathrm{neg}}<\frac{1}{2}4^n$ this implies that we only consider the $2n_{\mathrm{pos}}$ largest eigenvalues and discard the others. As already mentioned, in our implementation we chose $n_{\mathrm{pos}}=n_{\mathrm{neg}}=\frac{1}{2}4^n$.\footnote{We chose $n_{\mathrm{pos}}=n_{\mathrm{neg}}=1$ for $n=2$ and $n_{\mathrm{pos}}=n_{\mathrm{neg}}=8$ for $n=2$.} This way 
\begin{equation*}
    \Lambda_\mathcal{F} = \sum\limits_{i=1}^{n_{\mathrm{pos}}}{\tilde{\Lambda}_{\mathcal{G}_i}^+} - \sum\limits_{i=1}^{n_{\mathrm{neg}}}{\tilde{\Lambda}_{\mathcal{G}_i}^-}
\end{equation*}
is already fulfilled by the initial guess. Furthermore we also already fulfill that the $\gamma$-factor is
\begin{equation*}
     \sum_{i=1}^{n_{\mathrm{pos}}} a_i^+ + \sum_{i=1}^{n_{\mathrm{neg}}} a_i^- = f^* \, .
\end{equation*}
Therefore the only condition that is not yet fulfilled is the trace-preservingness of the $\tilde{\Lambda}_{\mathcal{G}_i}^{\pm}$.

\section{Heuristics for Stinespring algorithm}\label{app_heuristicts}
It is natural to ask whether we can prove the convergence of the Stinespring algorithm presented in~\Cref{sec_stinespring} or derive bounds on the approximation error. Unfortunately, it seems difficult to even prove convergence, as several heuristics were used. In this section we aim to provide a clear overview of these heuristics.

\paragraph{Existence of a low-rank channel decomposition.} It is not immediately clear if~\Cref{ass:bm_rank_solution} holds in practice or not. The fact that we observed a value of $\varepsilon>0$ to be necessary for good convergence could be an indicator that it does indeed not hold. Furthermore it is not clear what minimal values of $n_{\mathrm{pos}},n_{\mathrm{neg}}$ are required to reach this minimum.
  
\paragraph{Burer-Monteiro convergence.} In recent years, there has been an impressive progress in mathematically demonstrating the convergence and convergence speed of the Burer-Monteiro heuristic~\cite{Boumal2016, Cifuentes2019}. However, it is not clear if they can be applied to our setting.

\paragraph{Quality of the Stinespring dilation approximation.} Even under the assumption that there exists a solution of problem~(\ref{eq:rank_constrained_cd}) and that we can find it efficiently, the convergence speed of the Stinespring algorithm strongly depends on how well we can approximate an arbitrary quantum channel with the Stinespring dilation. In technical terms, we require the noise oracle $\mathcal{N}$ to be as closed to the identity as possible. It would be interesting for future work to prove the convergence speed and resulting $\gamma$-factor depending on some kind of fidelity measure of $\mathcal{N}$. 

\section{Simulation results for depolarizing noise}
\begin{figure}[!htb]
    \centering
    \includegraphics[scale=0.75]{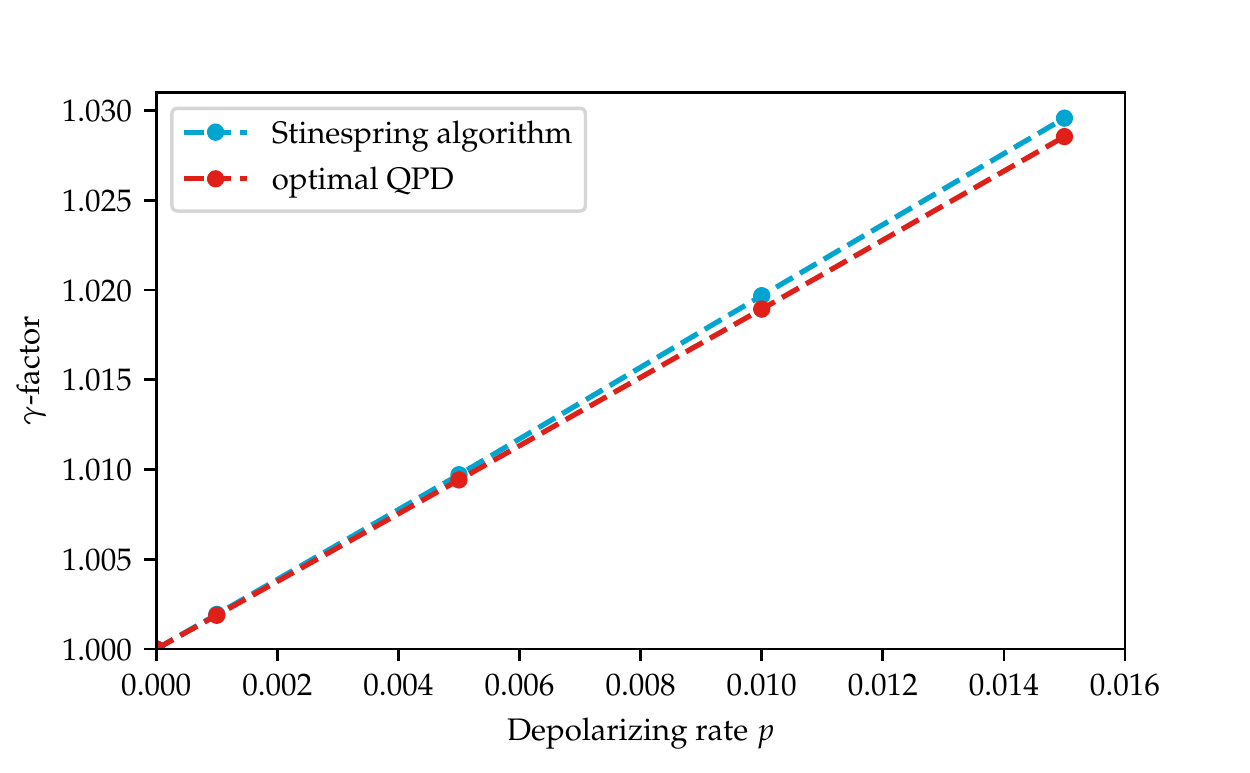} 
    \caption{Comparison of $\gamma$-factor obtained from Stinespring algorithm with optimal $\gamma$-factor. The depolarizing rate $p$ for two-qubit gates is varied. The depolarizing gate of single-qubit gates is assumed to be $0.1p$. The $\gamma$-factor of the optimal QPD is computed using the channel difference decomposition introduced in~\Cref{sec:rank_constrained_cd}.}
    \label{fig:depol_simulation}
\end{figure}

In case the gate errors consist of purely depolarizing noise, the optimal QPD to realize the inverse map is analytically known and its corresponding decomposition set only contains noisy Pauli operations~\cite{takagi20}.
So for practical considerations, there is not much use in applying the Stinespring algorithm to this specific type of noise.
However, it can still be interesting to run the Stinespring algorithm and compare the quality of the obtained QPD to the ideal one.
Indeed in~\cref{fig:depol_simulation}, we report that the Stinespring algorithm is able to recover a QPD that is almost as good as the optimal one.
The value of $\gamma-1$ only differs only by roughly 3\% to 4\% from the optimal value.
The simulations are performed for different values of the depolarizing error rate $p$ of two-qubit gates, where the depolarizing rate of single-qubit gates is assumed to be $0.1p$.

\section{Comparison with standard basis QPD}
\begin{figure}[!htb]
    \centering
    \subfloat[]{\includegraphics[width=0.49\textwidth]{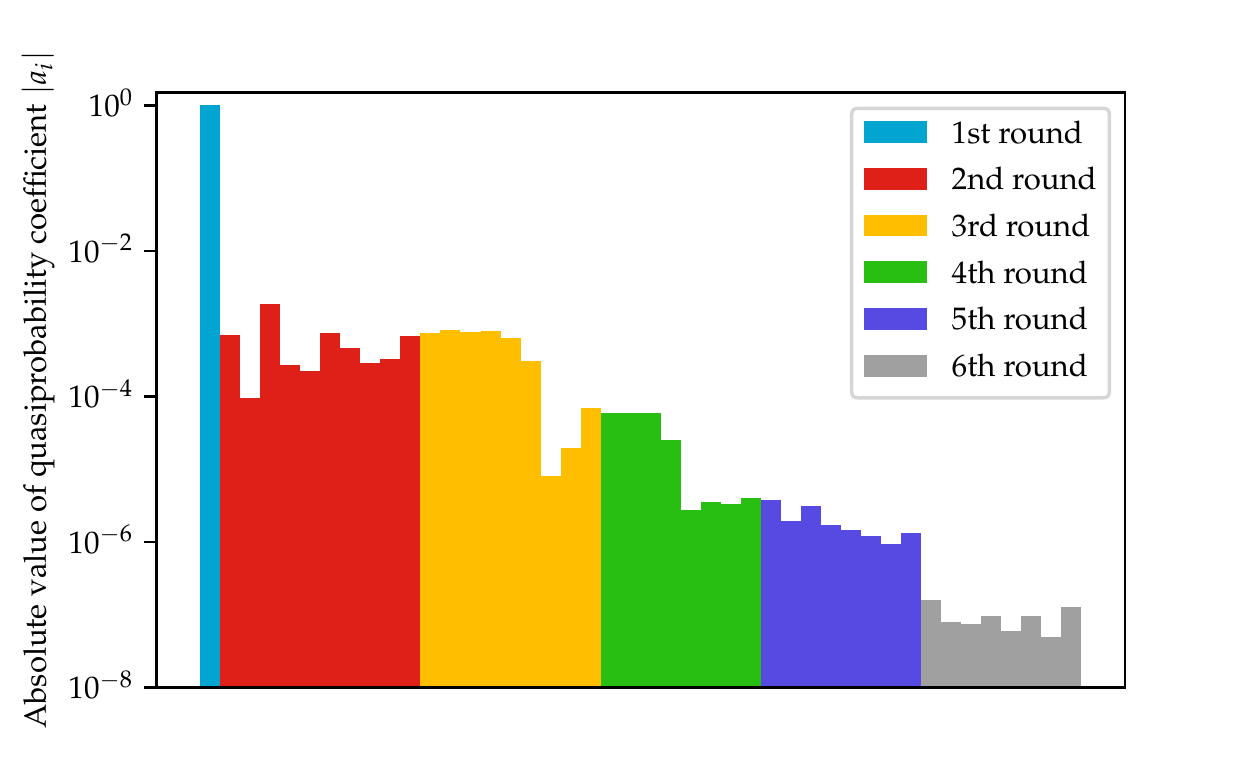}}
    \subfloat[]{\includegraphics[width=0.49\textwidth]{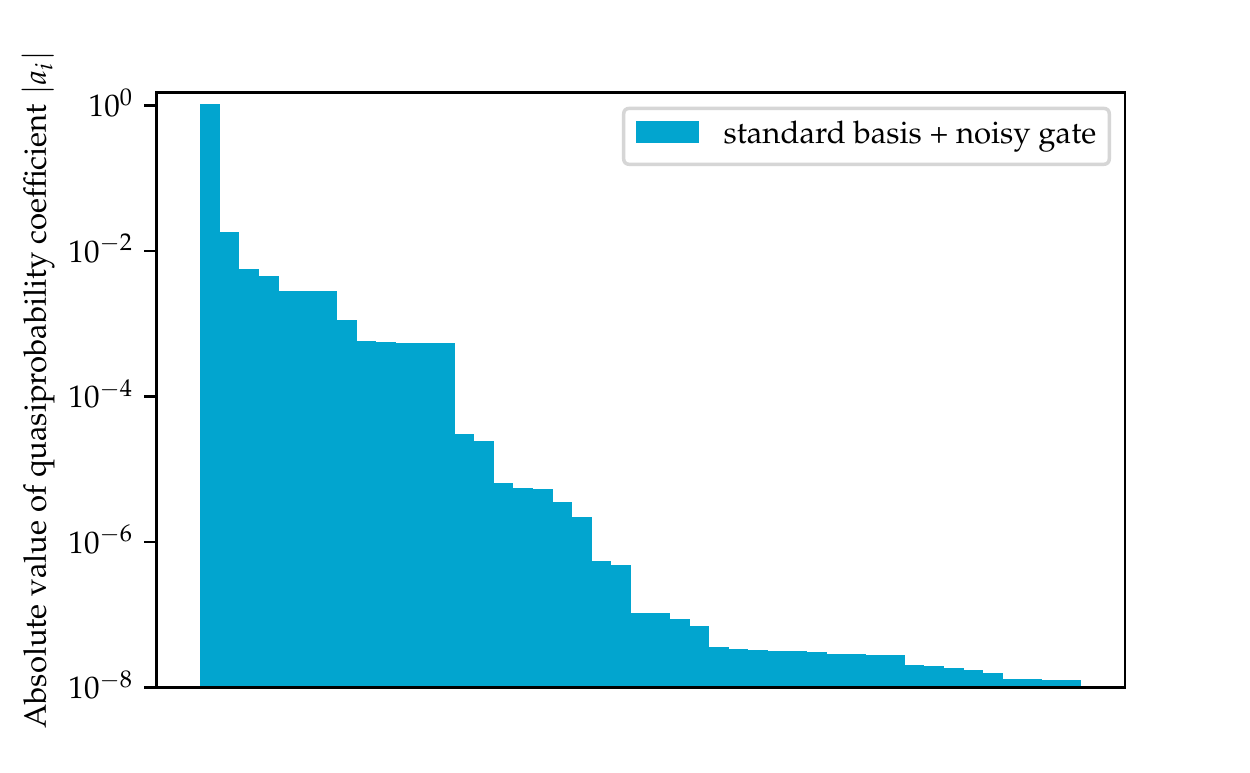}}
    \caption{Absolute value of quasiprobability coefficients obtained when decomposing the noisy $\CNOT$ gate discussed in~\cref{sec_simulationResults} into the decomposition sets (a) obtained form the Stinespring algorithm and (b) consisting of the standard basis and the noisy gate itself. Only quasiprobability coefficients with absolute value larger than $10^{-8}$ are displayed.}
    \label{fig_coeffs}
\end{figure}

The goal of this section is to provide further illustrations in order to provide some insight into how the decomposition set obtained from the Stinespring algorithm outperforms the decomposition set consisting of the standard basis.
For this purpose, we consider the simulation results obtained in~\Cref{sec_simulationResults} for the $\CNOT$ gate using the noise model of \texttt{ibmq\_melbourne}.
\Cref{fig_coeffs} (a) depicts the absolute value of the quasiprobability coefficients of the QPD corresponding to the decomposition set obtained from the Stinespring algorithm.
The colors indicate in which iteration of the Stinespring algorithm a specific element was added to the decomposition set.
It can be seen that the quasiprobability coefficients from later rounds have significantly lower contribution to the $\gamma$-factor than the earlier ones.
Similarly,~\Cref{fig_coeffs} (b) depicts the magnitude of the quasiprobability coefficients ordered in descending order when the ideal gate is decomposed into the decomposition set consisting of the standard basis as well as the noisy realisation of the gate.
Due to the larger number of operations which have a large quasiprobability coefficient, the resulting $\gamma$-factor is higher.
\Cref{tab_highestweights} depicts the 13 operations in the decomposition set which exhibit the largest magnitude of their quasiprobability weights.

\begin{table}[!htb]
    \centering
    \begin{tabular}{|l c|}
        \hline
        Operation & Quasiprobability coefficient \\
        \hline
        noisy gate & 1.03788 \\
        $\mathbb{1}\otimes\pi_Z$ & -0.02737 \\
        $X\otimes\pi_Z$ & 0.01855 \\
        $X\otimes Z$ & -0.00928 \\
        $X\otimes\pi_{XY}$ & -0.00891 \\
        $X\otimes\mathbb{1}$ & -0.00642 \\
        $\pi_{XY}\otimes Z$ & -0.00578 \\
        $\pi_Z\otimes Z$ & -0.00577 \\
        $\pi_{XY}\otimes\mathbb{1}$ & -0.00575 \\
        $\pi_Z\otimes\mathbb{1}$ & -0.00574 \\
        $R_x\otimes R_z$ & 0.00567 \\
        $\Pi_{YZ}\otimes Z$ & -0.00455 \\
        $\pi_{YZ}\otimes\mathbb{1}$ & 0.00455 \\
        \hline
    \end{tabular}
    \caption{13 elements in the decomposition set with highest magnitude quasiprobability coefficients. These table entries correspond to the 13 leftmost bars in~\Cref{fig_coeffs} (b).}
    \label{tab_highestweights}
\end{table}

\bibliographystyle{arxiv_no_month}
\bibliography{bibliofile}

\begin{thebibliography}{10}

\bibitem{Qiskit}
H.~Abraham et~al.
\newblock Qiskit: An open-source framework for quantum computing, 2019.
\newblock
  \texttt{\href{http://dx.doi.org/10.5281/zenodo.2562110}{DOI:\,10.5281/zenodo.2562110}}.

\bibitem{Agrawal2018}
A.~Agrawal, R.~Verschueren, S.~Diamond, and S.~Boyd.
\newblock A rewriting system for convex optimization problems.
\newblock {\em Journal of Control and Decision}, 5(1):42--60, 2018.
\newblock
  \texttt{\href{http://dx.doi.org/10.1080/23307706.2017.1397554}{DOI:\,10.1080/23307706.2017.1397554}}.

\bibitem{AB08}
D.~Aharonov and M.~Ben-Or.
\newblock Fault-tolerant quantum computation with constant error rate.
\newblock {\em SIAM J. Comput.}, 38(4):1207–1282, 2008.
\newblock
  \texttt{\href{http://dx.doi.org/10.1137/S0097539799359385}{DOI:\,10.1137/S0097539799359385}}.

\bibitem{Boumal2016}
N.~Boumal, V.~Voroninski, and A.~Bandeira.
\newblock The non-convex {B}urer-{M}onteiro approach works on smooth
  semidefinite programs.
\newblock In {\em Advances in Neural Information Processing Systems},
  volume~29, 2016.
\newblock Available online:
  \url{https://proceedings.neurips.cc/paper/2016/file/3de2334a314a7a72721f1f74a6cb4cee-Paper.pdf}.

\bibitem{Burer2003}
S.~Burer and R.~Monteiro.
\newblock A nonlinear programming algorithm for solving semidefinite programs
  via low-rank factorization.
\newblock {\em Mathematical Programming, Series B}, 95:329--357, 2003.
\newblock
  \texttt{\href{http://dx.doi.org/10.1007/s10107-002-0352-8}{DOI:\,10.1007/s10107-002-0352-8}}.

\bibitem{Burer2005}
S.~Burer and R.~Monteiro.
\newblock Local minima and convergence in low-rank semidefinite programming.
\newblock {\em Mathematical Programming}, 103:427--444, 2005.
\newblock
  \texttt{\href{http://dx.doi.org/10.1007/s10107-004-0564-1}{DOI:\,10.1007/s10107-004-0564-1}}.

\bibitem{cai21}
Z.~Cai.
\newblock Multi-exponential error extrapolation and combining error mitigation
  techniques for nisq applications.
\newblock {\em npj Quantum Information}, 7(1):80, 2021.
\newblock
  \texttt{\href{http://dx.doi.org/10.1038/s41534-021-00404-3}{DOI:\,10.1038/s41534-021-00404-3}}.

\bibitem{Cifuentes2019}
D.~Cifuentes.
\newblock On the {B}urer--{M}onteiro method for general semidefinite programs.
\newblock {\em Optimization Letters}, 2021.
\newblock
  \texttt{\href{http://dx.doi.org/10.1007/s11590-021-01705-4}{DOI:\,10.1007/s11590-021-01705-4}}.

\bibitem{Diamond2016}
S.~Diamond and S.~Boyd.
\newblock {CVXPY}: {A} {P}ython-embedded modeling language for convex
  optimization.
\newblock {\em Journal of Machine Learning Research}, 17(83):1--5, 2016.

\bibitem{Endo2018}
S.~Endo, S.~C. Benjamin, and Y.~Li.
\newblock Practical quantum error mitigation for near-future applications.
\newblock {\em Phys. Rev. X}, 8:031027, 2018.
\newblock
  \texttt{\href{http://dx.doi.org/10.1103/PhysRevX.8.031027}{DOI:\,10.1103/PhysRevX.8.031027}}.

\bibitem{JWW20}
J.~Jiang, K.~Wang, and X.~Wang.
\newblock Physical implementability of quantum maps and its application in
  error mitigation, 2020.
\newblock Available online: \url{https://arxiv.org/abs/2012.10959}.

\bibitem{Scipy}
E.~Jones et~al.
\newblock {SciPy}: Open source scientific tools for {Python}, 2001--.
\newblock Available online: \url{http://www.scipy.org/}.

\bibitem{Khatri2019}
S.~Khatri, R.~LaRose, A.~Poremba, L.~Cincio, A.~T. Sornborger, and P.~J. Coles.
\newblock Quantum-assisted quantum compiling.
\newblock {\em {Quantum}}, 3:140, 2019.
\newblock
  \texttt{\href{http://dx.doi.org/10.22331/q-2019-05-13-140}{DOI:\,10.22331/q-2019-05-13-140}}.

\bibitem{Li2017}
Y.~Li and S.~C. Benjamin.
\newblock Efficient variational quantum simulator incorporating active error
  minimization.
\newblock {\em Phys. Rev. X}, 7:021050, 2017.
\newblock
  \texttt{\href{http://dx.doi.org/10.1103/PhysRevX.7.021050}{DOI:\,10.1103/PhysRevX.7.021050}}.

\bibitem{Maclaurin2015}
D.~Maclaurin, D.~Duvenaud, and R.~P. Adams.
\newblock Autograd: Effortless gradients in {NumPy}.
\newblock {\em ICML 2015 AutoML Workshop}, 2015.
\newblock Available online:
  \url{https://indico.lal.in2p3.fr/event/2914/session/1/contribution/6/3/material/paper/0.pdf}.

\bibitem{McClean2017}
J.~R. McClean, M.~E. Kimchi-Schwartz, J.~Carter, and W.~A. de~Jong.
\newblock Hybrid quantum-classical hierarchy for mitigation of decoherence and
  determination of excited states.
\newblock {\em Phys. Rev. A}, 95:042308, 2017.
\newblock
  \texttt{\href{http://dx.doi.org/10.1103/PhysRevA.95.042308}{DOI:\,10.1103/PhysRevA.95.042308}}.

\bibitem{Mosek}
{MOSEK ApS}.
\newblock {\em MOSEK Optimizer API for Python 9.2.8}, 2020.
\newblock Available online:
  \url{https://docs.mosek.com/9.2/pythonapi/index.html}.

\bibitem{Otten2018}
M.~Otten and S.~Gray.
\newblock Accounting for errors in quantum algorithms via individual error
  reduction.
\newblock {\em npj Quantum Information}, 5, 2018.
\newblock
  \texttt{\href{http://dx.doi.org/10.1038/s41534-019-0125-3}{DOI:\,10.1038/s41534-019-0125-3}}.

\bibitem{Piv_masterThesis}
C.~Piveteau.
\newblock Advanced methods for quasiprobabilistic quantum error mitigation.
\newblock
  \texttt{\href{http://dx.doi.org/10.3929/ethz-b-000504508}{DOI:\,10.3929/ethz-b-000504508}}.
\newblock Master thesis, ETH Zurich, September 2020.

\bibitem{Preskill2018}
J.~Preskill.
\newblock Quantum computing in the {NISQ} era and beyond.
\newblock {\em Quantum}, 2:79, 2018.
\newblock
  \texttt{\href{http://dx.doi.org/10.22331/q-2018-08-06-79}{DOI:\,10.22331/q-2018-08-06-79}}.

\bibitem{Armands2020}
A.~Strikis, D.~Qin, Y.~Chen, S.~C. Benjamin, and Y.~Li.
\newblock Learning-based quantum error mitigation, 2020.
\newblock Available online: \url{https://arxiv.org/abs/2005.07601}.

\bibitem{takagi20}
R.~Takagi.
\newblock Optimal resource cost for error mitigation, 2020.
\newblock Available online: \url{https://arxiv.org/abs/2006.12509}.

\bibitem{Temme2017}
K.~Temme, S.~Bravyi, and J.~M. Gambetta.
\newblock Error mitigation for short-depth quantum circuits.
\newblock {\em Physical Review Letters}, 119(18), 2017.
\newblock
  \texttt{\href{http://dx.doi.org/10.1103/physrevlett.119.180509}{DOI:\,10.1103/physrevlett.119.180509}}.

\bibitem{Watrous2013}
J.~Watrous.
\newblock Simpler semidefinite programs for completely bounded norms.
\newblock {\em Chicago Journal of Theoretical Computer Science}, 2013(8), 2013.
\newblock
  \texttt{\href{http://dx.doi.org/10.4086/cjtcs.2013.008}{DOI:\,10.4086/cjtcs.2013.008}}.

\end{thebibliography}

\end{document}